\documentclass[iop]{emulateapj}

\usepackage{natbib}
\usepackage{epsfig}
\usepackage{comment}
\usepackage{float}
\usepackage{hyperref}
\usepackage{amssymb}
\usepackage{color}
\usepackage[normalem]{ulem}

\newcommand{\na}{Nature}

\newcommand{\fhtwo}{f_{\rm{H}_2}}
\newcommand{\tff}{t_{ff}}
\newcommand{\gadget}{{\small GADGET-3}\ }
\newcommand{\msun}{\rm{M}_{\odot}}
\newcommand{\mstar}{M_{\star}}

\newcommand{\mtot}{M_{\rm{tot},200}}
\newcommand{\mdm}{M_{\rm{DM}}}
\newcommand{\htwo}{\rm{H}_{2}}
\newcommand{\Om}{\Omega_{\rm m}}
\newcommand{\Ol}{\Omega_{\Lambda}}
\newcommand{\Ob}{\Omega_{\rm b}}

\newcommand{\Zsun}{\rm{Z}_{\odot}}
\newcommand{\Msun}{\rm{M}_{\odot}}
\newcommand{\mgas}{M_{\rm{gas}}}
\newcommand{\sigmasfr}{\dot{\Sigma}_\star}
\newcommand{\sigmaHI}{\Sigma_{\rm{HI}}}
\newcommand{\sigmahtwo}{\Sigma_{\rm{H2}}}
\newcommand{\HI}{H{\sc i}}
\newcommand{\fgas}{f_{\rm gas}}
\newcommand{\fHI}{f(N_{\rm HI})}
\newcommand{\NHI}{N_{\rm HI}}
\newcommand{\fb}{f_{\rm b}}

\newcommand{\sigmasob}{\Sigma_{\rm{Sob}}}
\newcommand{\sigmagrid}{\Sigma_{\rm{grid}}}

\newcommand {\apgt} {\ {\raise-.5ex\hbox{$\buildrel>\over\sim$}}\ }
\newcommand {\aplt} {\ {\raise-.5ex\hbox{$\buildrel<\over\sim$}}\ }

\shorttitle{Molecular hydrogen regulated star formation in cosmological SPH simulations}
\shortauthors{Robert Thompson, Kentaro Nagamine, Jason Jaacks, Jun-Hwan Choi}

\begin{document}
\title{Molecular hydrogen regulated star formation in cosmological SPH simulations}

\author{Robert Thompson\altaffilmark{1}, Kentaro Nagamine\altaffilmark{1,2}, Jason Jaacks\altaffilmark{1}, Jun-Hwan Choi\altaffilmark{3}}

\altaffiltext{1}{Department of Physics \& Astronomy, University of Nevada Las Vegas, 4505 S. Maryland Pkwy, Box 454002, Las Vegas, NV 89154-4002 U.S.A.; \quad Email: rthompson@physics.unlv.edu} 
\altaffiltext{2}{Department of Earth and Space Science, Graduate School of Science, Osaka University, 1-1 Machikaneyama-cho, Toyonaka, Osaka, 560-0043, Japan}
\altaffiltext{3}{Department of Physics \& Astronomy, University of Kentucky, Lexington, KY}








\begin{abstract}
Some observations have shown that star formation (SF) correlates tightly with the presence of molecular hydrogen ($\htwo$), 
therefore it would be important to investigate its implication on galaxy formation in a cosmological context. 
In the present work, we implement a sub-grid model (hereafter $\htwo$-SF model) that tracks the $\htwo$ mass fraction within our cosmological smoothed particle hydrodynamics (SPH) code \gadget using an equilibrium analytic model by Krumholz et al.
This model allows us to regulate the star formation in our simulation by the local abundance of $\htwo$ rather than the total cold gas density, which naturally introduces the dependence of star formation on metallicity. 
We investigate the implications of the $\htwo$-SF model on galaxy population properties, such as the stellar-to-halo mass ratio (SHMR), baryon fraction, cosmic star formation rate density (SFRD), galaxy specific SFR, galaxy stellar mass functions (GSMF), and the Kennicutt-Schmidt (KS) relationship. 
The advantage of our work over the previous ones is having a large sample of simulated galaxies in a cosmological volume 
from high-redshift to $z=0$.  
We find that low-mass halos with $\mdm <10^{10.5}\msun$ are less efficient in producing stars in the $\htwo$-SF model at $z\geq6$, which brings the simulations to a better agreement with the observational estimates of SHMR and GSMF at the low-mass end. 
This is particularly evident by a reduction in the number of low-mass galaxies at $\mstar \leq10^{8}\msun$ in the GSMF.
The overall SFRD is also reduced at high-$z$ in the $\htwo$ run, which results in slightly higher SFRD at low-redshift due to 
more abundant gas available for star formation at later times. 
This new $\htwo$ model is able to reproduce the empirical KS relationship at $z=0$ naturally without the need for setting its normalization by hand, and overall it seems to have more advantages than the previous pressure-based SF model.  
\end{abstract}


\section{Introduction}
\label{sec:intro}

Properly modeling star formation (SF) and feedback within simulations of galaxy formation is one of the holy grails for computational astrophysicists.  Unfortunately, current cosmological simulations of galaxy formation lack the spatial and mass resolutions to properly resolve the small scale processes that govern star formation within the interstellar medium (ISM).  This computational restriction gives rise to the need for sub-grid models that can accurately describe global properties of the ISM.  Simulation results can vary drastically depending on the details adopted for such sub-grid models and their feedback prescriptions.  It is for this reason that these sub-grid models rely heavily on observed empirical star formation models.

The most well-known empirical star formation relation is the \citet{Schmidt59} and \citet{Kennicutt98} relationship, which relates the density (or surface density) of star formation to the gas density (or surface density), respectively. 
For numerical simulations of galaxy formation, the Schmidt relationship is easier to implement \citep[e.g.,][]{Katz92, Cen92},
however, observationally the Kennicutt relationship is easier to measure because observations are done in the projected two  dimensional plane. 
In the present work, we are implementing the Schmidt relationship as part of our SF model, but when comparing to the observations we use the Kennicutt relationship, hence referring to them collectively as the Kennicutt-Schmidt (KS) relationship. 

Recent observational evidence has suggested that star formation is more tightly correlated with the presence of molecular hydrogen ($\htwo$), rather than neutral atomic (\HI) hydrogen \citep{Wong02,Kennicutt07,Leroy08,Bigiel08,Bolatto11}.  
In particular, \citet{Bigiel08} studied the KS relation for a sample of nearby galaxies, and found little to no correlation between $\sigmaHI$ and $\sigmasfr$, whereas $\sigmahtwo$ was found to correlate strongly with $\sigmasfr$. 
\citet{Bolatto11} used the \emph{Spitzer} observations of dust continuum in the low-metallicity SMC to calculate $\htwo$ surface densities without the need for a CO luminosity conversion factor.  
Their findings suggested that $\htwo$ can be used to infer star formation activity even in low-metallicity galaxies.


\begin{deluxetable*}{cccccccccc}
\tablecolumns{10}
\tablewidth{0pc}
\tablehead{\multicolumn{1}{c}{Run Name} &
	\colhead{Box Size} &
	\colhead{Particle Count} &
	\colhead{$m_{\rm{dm}}$} &
	\colhead{$m_{\rm{gas}}$} &
	\colhead{$\epsilon$} &
	\colhead{$z_{\rm{end}}$} &
	\colhead{$z_{\rm{end}}$} &
	\colhead{OTUV} &
	\colhead{OTUV} \\
	\colhead{ \ } &
	\colhead{[$h^{-1}$ Mpc]} &
	\colhead{DM \& Gas} &
	\colhead{[$h^{-1}$ $\Msun$]} &
	\colhead{[$h^{-1}$ $\Msun$]} &
	\colhead{[$h^{-1}$ kpc]} &
	\colhead{$\htwo$} &
	\colhead{Fiducial} &
	\colhead{$\htwo$} &
	\colhead{Fiducial}
}
\startdata
N144L10	&	10.00	&	$2\times144^3$	&	$2.01\times10^7$	&	$4.09\times10^6$	&	2.77	& 3.00 & 3.00 & Y & Y\\
N500L34 	&	33.75	&	$2\times500^3$	&	$1.84\times10^7$	&	$3.76\times10^6$	&	2.70	& 3.00 & - & Y & -\\
N600L10 	&	10.00	&	$2\times600^3$	&	$2.78\times10^5$	& 	$5.65\times10^4$	&	0.67	& 6.00 & - & Y & -\\
\\
\sidehead{Production Runs}
N400L10	&	10.00	&	$2\times400^3$	&	$9.37\times10^5$	&	$1.91\times10^5$	&	1.00  & 6.00  & 5.50 & Y & N\\
N400L34	&	33.75	&	$2\times400^3$	&	$3.60\times10^7$	&	$7.34\times10^6$	&	3.38  & 3.00  & 1.00 & Y & N\\
N600L100	&	100.00	&	$2\times600^3$	&	$2.78\times10^8$	&	$5.65\times10^7$	&	4.30   & 0.00 & 0.00 & Y & N\\
\enddata
\tablecomments{
Simulation parameters used in this work.  The first three simulations were used to perform tests of the $\htwo$ model and resolution study (Section~\ref{sec:resdepend}).  The second set of three simulations are the main production runs used to compare with previous SF models.
The quantities m$_{\rm{dm}}$ \& m$_{\rm{gas}}$ are the particle masses of dark matter and gas particles, $\epsilon$ is the comoving gravitational softening length, and $z_{\rm{end}}$ is the ending redshift of each simulation.
The $\htwo$ simulations (along with N144L10 Fiducial) use an optically-thick ultra-violet threshold or `OTUV' \citep[see Section~\ref{sec:phase};][]{Nagamine10}.
}
\label{table:sims}
\end{deluxetable*}

Driven by these observational findings, new models have been developed relating SFRs directly to the abundance of $\htwo$.
Some are in the form of analytic models \citep{Fu10,Krumholz08,Krumholz09,McKee10,KrumholzDekel12}, 
while others in the form of non-equilibrium, fully time-dependent calculations \citep{Gnedin09, Feldmann11,MacLow12}. However, many of these models have been restricted to single isolated galaxies or cosmological zoom-in simulations of a very small sample of galaxies due to the expensive computational cost of full cosmological simulations. 

Recently, both semi-analytic and non-equilibrium $\htwo$ calculations have been implemented into full cosmological simulations.
\citet{Kuhlen12,Kuhlen13} implemented the analytic model of \citet{Krumholz08, Krumholz09} and \citet{McKee10} 
in the adaptive-mesh-refinement code {\small Enzo} \citep{ENZO1,ENZO2,ENZO3}
to study how $\htwo$-based star formation affected dwarf galaxies at $z>2.5$. 
Both their previous model and the new $\htwo$ model were able to reproduce many of the observational results pertaining to the KS relation.  The advantage they found within the $\htwo$ model was that it reduced the number of free parameters, and that star formation was quenched in dwarf galaxies from the onset without the need to artificially enhance stellar feedback.
\citet{Christensen12} implemented the non-equilibrium, fully time-dependent model of \citet{Gnedin09} into their cosmological SPH code {\small GASOLINE} \citep{GASOLINE} 
in order to study the effects of $\htwo$-SF model on a dwarf galaxy down to $z=0$.
They found that the inclusion of $\htwo$ resulted in a greater baryonic mass in the disk, making it brighter, bluer, and more gas rich at $z=0$ than the same galaxy formed without the inclusion of $\htwo$.  
They also found that with $\htwo$-SF model, there was more star formation at late times.
Since SF occurs on scales much smaller than the galaxy as a whole, most of the previous studies employed very high-resolution simulations to verify the significance of the $\htwo$-SF model.  Owing to the computational limitation, these studies did not investigate the effect of the improved SF model on the statistical properties of galaxies that requires large cosmological volumes.

While there are other models of star formation based on, for example, supersonic turbulence in the ISM \citep[e.g.][]{Krumholz05,Mckee07,Padoan11,Hennebelle11,Kritsuk11,Renaud12}, 
it is still worthwhile to explore an implementation of the $\htwo$-SF model as well, and investigate its implications on the overall galaxy populations. 
The purpose of this paper is not to decide which process triggers the star formation (i.e., supersonic turbulence or molecules), as our simulations have neither the resolution nor detailed dust physics to address the issue. 
In this paper, we limit ourselves to examining the effects of the $\htwo$-SF model on galaxy formation, 
and we defer the implementation of the turbulence-based SF model to the future. 
We would like to stress that the goal of this paper is to change the SF prescription only, leaving the remaining physics unchanged from previous models.  This will allow us to isolate the effects of the new H2-SF prescription.
Some minor modifications were made in addition to the new SF model but were found to have little impact on the results of this paper;  these changes will be discussed in Section~\ref{sec:sims}.

There is another practical reason related to observations to study the $\htwo$-SF model in cosmological simulations.
Many of the earlier works based on a cold dark matter (CDM) model have predicted very steep faint-end slopes of the mass/luminosity functions at high-redshift \citep[e.g.,][]{Nagamine04groupfinder, Night06, LoFaro09, Trenti10, Salvaterra11, Finlator11, Jaacks12}, 
and suggested that these low-mass galaxies are responsible for reionizing the Universe at $z\ge 6$.   However, the observational estimates yield slightly shallower faint-end slopes, and if they are not affected by the magnitude limit very much,  the simulations need to consider processes that would decrease the number of low-mass galaxies, especially at high redshift.  One of such candidate processes would be $\htwo$-based star formation, and \citet{Jaacks12} for example have speculated that the $\htwo$-SF model may reduce the discrepancy in galaxy stellar mass function (GSMF) at the low-mass end. 

This paper is organized as follows.  In Section~\ref{sec:sims} we discuss simulation parameters, SF models, and basic results.  Section~\ref{sec:simresults} contains our findings for galaxy populations.  The results of SHMR, cosmic SFRD, GSMF, and KS relation are presented along with resolution studies.
Lastly in Section~\ref{sec:summary} we summarize our results and discuss future prospects.


\section{Simulations \& basic results}
\label{sec:sims}

We use a modified version of the \gadget cosmological SPH code \citep[originally described in][]{Springel05}.  
Our conventional code includes radiative cooling by H, He, and metals \citep{Choi09}, 
heating by a uniform UV background (UVB) of a modified \citet{Haardt96} spectrum \citep{Faucher09},  
supernova (SN) feedback, 
and the Multi-component Variable Velocity (MVV) wind model \citep{Choi11}. 
We also use the sub-resolution multiphase ISM model \citep{Springel03}, where 
the high-density ISM is pictured to be a two-phase fluid consisting of cold clouds in pressure equilibrium with a hot ambient phase.
Hydrodynamic forces for multiphase particles are calculated only for the hot phase.
The cold phase on the other hand provides material for star formation, is subject to gravity, adds inertia, and participates in mass \& energy exchange with the hot phase at the sub-particle level. 
The primary purpose of this work is to improve upon the SF models implemented within this code.  
Previous SF model implementations are also discussed in upcoming sections.

The same set of initial conditions (ICs) used by \citet{Choi11} and \citet{Jaacks12} are employed in this study to allow for cross comparison. 
Three primary simulations were run consisting of 2$\times$400$^3$ or 2$\times$600$^3$ particles of gas and dark matter.  Comoving box sizes of $10h^{-1}$Mpc, $34h^{-1}$Mpc, \& $100h^{-1}$Mpc are used to capture a range of halo masses.  These runs will be referred to by their particle count followed by the length of their box: N400L10, N400L34, \& N600L100. 
The other three runs (N144L10, N500L34, N600L10) where used mainly for resolution tests of the $\htwo$ model, and 
N500L34 \& N600L10 runs were performed only for the $\htwo$ model due to expensive computational cost. 
The ICs were generated using cosmological parameters consistent with the WMAP best-fit values \citep{WMAP7}: $\Om=0.26$, $\Ol=0.74$, $\Ob=0.044$, $H_0/100=0.72$, $n_s=0.96$, and $\sigma_8=0.80$.
The simulation parameter values are summarized in Table~\ref{table:sims}.
The runs with smaller box sizes are stopped at earlier times, because they do not include longer wavelength perturbations. 

There are three additional differences between what we will refer to as the `Fiducial' runs \citep{Choi11,Jaacks12} and the `$\htwo$' runs.  First, we use an updated model of UVB radiation in the $\htwo$ runs, as will be discussed in Section~\ref{sec:metalfloor}.
The Fiducial run uses an older model wherein the UVB turns on at $z=6.08$ to mimic the sudden reionization of the Universe; the UVB of the updated model turns on at $z=10.75$, given the more recent WMAP results on early reionization of the Universe. 
For the majority of our comparisons, this change has little impact.  
Secondly, our $\htwo$ runs use an optically-thick ultra-violet threshold or `OTUV' \citep{Nagamine10} which will be discussed in Section~\ref{sec:phase}.  Most comparisons presented in this paper are not effected by this; the column density distribution presented in Section~\ref{sec:neutralhydrogen} and Figure~\ref{fig:neutralcolumn} however, is strongly effected by this change.  Our low resolution (N144L10) Fiducial run uses the OTUV threshold. 
Lastly, the Fiducial run uses the \citet{Salpeter55} initial mass function (IMF), while our new runs use the \citet{Chabrier03} IMF.  The choice of the IMF is reflected in the value of gas recycling fraction parameter $\beta$ in the SF relation in our simulation. 
We have also verified that this has little impact on the results presented in this paper.

\subsection{H based SF models}
\label{sec:previousSF}
\subsubsection{SH model}
\label{sec:SHmodel}

\citet[][SH model]{Springel03} describes the hybrid multiphase model for SF, originally implemented in
the {\small GADGET} code. 
In this model, the SFR is determined by 
\begin{equation}
\dot{\rho}_* = (1-\beta) \frac{\rho_{cold}}{t_*},
\label{eq:springelsfr}
\end{equation}
where $\rho_{cold}$ is the density of cold gas available to form stars, and $\beta$ is the fraction of stars instantaneously destroyed as supernova, determined by the stellar IMF. 
The parameter $t_*$ is the SF time-scale which is taken to be proportional to the local dynamical time of the gas:
\begin{equation}
t_* = t_0^* \left(\frac{\rho}{\rho_{th}}\right)^{-1/2},
\label{eq:springeltff}
\end{equation}
where $\rho_{th}$ is a density threshold, above which 
the gas is assumed to develop a multiphase structure and form stars. 
The parameter $t_0^*$ controls the normalization of the \citet{Kennicutt98} relation:
\begin{equation}
\Sigma_{\rm{SFR}}=
\left\{ \begin{array}{lcl}
0 & \mbox{if} & \Sigma_{\rm{gas}} < \Sigma_{th} \\
A\left(\Sigma_{\rm{gas}}/1\rm{M}_{\odot} \rm{pc}^{-2}\right)^n & \mbox{if} & \Sigma_{\rm{gas}} > \Sigma_{th},
\end{array}\right.
\label{eq:kslaw}
\end{equation}
where $\Sigma_{th}$ is the SF threshold surface density.
Observations suggest 
$A=2.5\pm0.7 \times 10^{-4} \rm{M}_{\odot} \rm{yr}^{-1} \rm{kpc}^{-2}$, $n=1.4\pm0.15$, and $\Sigma_{th}\sim10\rm{M}_{\odot}\rm{pc}^{-2}$ \citep{Kennicutt98}.
The cutoff in $\Sigma_{\rm{SFR}}$ is controlled by $\rho_{th}$, which indirectly regulates $\Sigma_{th}$.
See \citet{Springel03} for a description of how $\rho_{th}$ is calculated self-consistently for this model.

Our simulations deal with three dimensional densities (i.e. Equation~\ref{eq:springelsfr}) rather than the two dimensional surface densities described by the KS-relation.  It is not obvious then that Eq.~(\ref{eq:springelsfr}) would be able to reproduce the observed KS-relation given by Eq.~(\ref{eq:kslaw}).
Previous simulations \citep{Springel03,Nagamine04sf} were able to demonstrate that the observed relation could indeed be reproduced using $t_0^*=2.1$\,Gyr, which resulted in a threshold value of $\rho_{th}=0.35\,h^{2}\rm{cm}^{-3}$.  
However, \citet{Nagamine04sf} and \citet{Choi10} showed that using this value of $\rho_{th}$ results in an overprediction of $\Sigma_{\rm{SFR}}$ at low column densities (N$_{\rm{HI}}$$\leq10^{20.5}\rm{cm}^{-2}$).
This over-prediction, coupled with the fact that this model produces stars from cold atomic gas rather than molecular, leads to the necessity for improvement in the sub-grid SF model, which we describe in the following sections.

\subsubsection{Pressure model}
\label{sec:pressuremodel}

Previous SF models assumed that the exponents of the Kennicutt and Schmidt relationships are equal.
This is only true if the galactic disk scale-height is constant, or the equation of state behaves as $P\propto \rho^2$ \citep{Schaye08}.
Arguing that these assumptions are unnecessary and often incorrect, \citet{Schaye08} formulated a fully analytic conversion from the 2D KS-relation $\left(\Sigma_{\rm{gas}}\right)$ to a 3D Schmidt-relation ($\rho_{\rm{gas}}$). 
They proposed that the density of a self-gravitating disk will fluctuate on the local Jeans scale, leading to the scale-height also being on the order of the local Jeans scale.  This in turn leads to the gas column density being on the order of the `Jeans column density': 
\begin{equation}
\label{eq:sigma_j}
\Sigma_{\rm{gas}}\sim\Sigma_{\rm{gas,J}}\equiv\rho_{\rm{gas}}L_{\rm{J}}=\left(\frac{\gamma}{G}\right)^{1/2} \left(f_gP_{tot}\right)^{1/2},
\end{equation}
where $L_{\rm{J}}=c_s/\sqrt{G\rho_{tot}}$ is the Jeans length, $c_s=\sqrt{\gamma P_{tot}/\rho_{\rm{gas}}}$ is the local sound speed, $\gamma$ is the ratio of specific heats, $G$ is the gravitational constant, $f_g$ is the mass fraction within the scale-height of the gas, and $P_{tot}$  is the mid-plane pressure.
The SFR in this model is also described by Equation~(\ref{eq:springelsfr}); the difference comes in the formulation of $t_0^*$ which is derived by combining Equations~(\ref{eq:kslaw}) \& (\ref{eq:sigma_j}):
\begin{equation}
t_0^* = \frac{\Sigma_{\rm{gas}}}{\Sigma_{\rm{SFR}}} = A^{-1} \left(1\, \msun \, \rm{pc}^{-2}\right)^n \left(\frac{\gamma}{G} \, f_{\rm{g}} \, P_{tot}\right)^{(1-n)/2}. 
\end{equation}
\citet{Schaye08} claim that their analytical conversion renders their parameters `untweakable'. 
Adopting $n=1.4$ \& $\gamma=5/3$, \citet{Choi10} implemented this model within our \gadget code with some minor modifications.  It was found to reduce the overprediction of $\Sigma_{\rm{SFR}}$ at low column densities and was in good agreement with the observed KS-relation.  
It also reduced the SFR in low-density regions, causing a suppression of early star formation, which in turn shifted the peak of the cosmic SFRD to lower redshifts in better agreement with observational estimates. 

The disadvantage of this model is that we are still imposing the KS relation onto our SF prescription.  In an ideal situation the model would naturally reproduce the observed KS relation without such impositions.
Simulations and data from previous work based off of the \citet{Schaye08} Pressure SF model \citep{Choi10,Choi11,Jaacks12} will serve as our Fiducial runs for comparison.

\subsection{$\htwo$ regulated star formation}
\label{sec:H2regsf}

If star formation really requires molecular gas, then tracking the $\htwo$ gas fraction and basing our SF prescription on it would make for a more realistic sub-grid model. 
There are currently two primary ways in which the $\htwo$ mass fraction can be tracked in simulations.
The first is a non-equilibrium model which calculates the $\htwo$ fraction via a fully time-dependent chemistry and radiative transfer calculation as was done by \citet{Gnedin09} and \citet{Christensen12}.  
This approach accurately calculates the instantaneous $\htwo$ gas fraction, but can be computationally expensive.
The second is a analytical approach developed by \citet{Krumholz08,Krumholz09} and \citet{McKee10} (hereafter KMT model), which calculates an equilibrium $\htwo$ fraction assuming a formation-dissociation balance.  

These two methods were directly compared in \citet{KrumholzGnedin11}; the second method was found to be a viable and nearly cost free alternative to the computationally expensive first option at metallicities $Z\geq 10^{-2} \Zsun$, where $\Zsun$ is the solar metallicity.
At metallicities $<10^{-2}\Zsun$ the KMT model was found to over-predict the fractional $\htwo$ abundance due to the neglect of time-dependent effects.
\citet{Krumholz12} however, argues that the equilibrium $\htwo$ fraction rather than the instantaneous one correlates more with gas temperature.
He argued that the thermal timescale of gas is much shorter than the chemical timescale, 
which means that low metallicity clouds should cool via collisional de-excitation and form stars faster than they can fully convert all of their atomic gas to molecular.
\citet{Glover12} came to a similar conclusion after investigating the CO-to-$\htwo$ conversion factors within low metallicity molecular clouds.
This suggests that the fractional $\htwo$ abundance calculated by the KMT model may more accurately predict the amount of material available to form stars in low metallicity gas.
Due to the computational simplicity we choose to adopt the KMT model to track $\htwo$ within our simulations, and examine its impact on cosmological galaxy formation. 

\subsubsection{KMT model}
\label{sec:H2model}

\citet{Krumholz08,Krumholz09} and \citet{McKee10} developed an analytic model for tracking $\htwo$ mass fraction.  They used a Str$\ddot{\rm{o}}$mgren-type analysis, starting with a spherical gas cloud immersed in a uniform, isotropic Lyman-Werner band radiation field.  Assuming that the cloud is in a steady state, they proceeded to solve the radiative transfer and $\htwo$ formation-dissociation balance equations.  After some approximations, they found a solution 
\begin{equation}
\fhtwo  \equiv \frac{\rho_{\rm{H}_2}}{\rho_{\rm{HI}}} \approx 1-\frac{3}{4}\left(\frac{s}{1+0.25s}\right),
\label{eq:fh2calc}
\end{equation}
where $\fhtwo$ is the $\htwo$ mass fraction relative to the neutral hydrogen gas.  
At such high densities where molecular gas may form, the hydrogen gas would be mostly neutral within our multiphase ISM model, hence the neutral hydrogen (\HI) in the denominator above (see also Section~\ref{sec:neutralassumption}). 
The parameter $s$ is essentially the size of the atomic-molecular complex given by
\begin{equation}
s\equiv\frac{\tau_R}{\tau_c}\xi_d= \frac{\ln\left(1+0.6\chi+0.01\chi^2\right)}{0.6\tau_c},
\label{eq:sval1}
\end{equation}
where $\tau_R$ is the dust optical-depth of the atomic-molecular complex, $\tau_c$ is the mean optical depth, and $\xi_d$ is the characteristic radius (in units of the cloud radius) at which the transition from dust-dominated to molecular-dominated absorption occurs.
The variable $\chi$ can be thought of as an estimation of the local radiation field given by
\begin{equation}
\chi=71\left(\frac{\sigma_{d,-21}}{\mathcal{R}_{-16.5}}\right)\frac{G_0'}{n_{\rm{H}}}.
\label{eq:chi1}
\end{equation}
Here $\sigma_{d,-21}$ is the dust cross section per H nucleus to 1000$\rm{\AA}$ radiation normalized to a value of 10$^{-21}$cm$^{2}$, $\mathcal{R}_{-16.5}$ is the rate coefficient for $\htwo$ formation on dust grains normalized to the Milky Way value of 10$^{-16.5}$cm$^{-3}$s$^{-1}$ \citep{Wolfire08},
$G_0'$ is the ambient UV radiation field intensity normalized to the \citet{Draine78} value for the Milky Way,
and $n_{\rm{H}}$ is the volume density of H nuclei.

At this point $\fhtwo$ depends only on $\tau_c$ \& $\chi$.  In order to calculate the dust optical depth ($\tau_c$), we first must calculate the local \HI\ column density $\left(\Sigma_{\rm{HI}}\right)$.
Given that we cannot resolve galactic disks within our cosmological simulations, we must use some approximation to estimate this value for individual SPH particles.
To do this we employ the Sobolev-like approximation \citep[e.g.][]{Gnedin09,KrumholzGnedin11}:
\begin{equation}
\Sigma_{\rm{HI}}\approx \rho_{\rm{HI}}\times h,
\label{eq:sobolev}
\end{equation}
where $h$ is the local scale height estimated by
\begin{equation}
h\approx \frac{\rho_{\rm{HI}}}{|\nabla \rho_{\rm{HI}}|}.
\label{eq:h}
\end{equation}
To test this approximation we calculate the column density along the z-axis of a representative halo within our simulation on a uniform grid with a cell size of $\epsilon^2$ (hereafter $\sigmagrid$; a detailed description of this procedure can be found in \citet{Nagamine04column}).  
When we examine the column density of each particle calculated using Eq. ~(\ref{eq:sobolev}) (hereafter $\sigmasob$), 
we find that the majority of the particles have $\sigmasob$ that are lower than $\sigmagrid$, but some are found to be higher than $\sigmagrid$. 
On the other hand, when we take the mass-weighted average of each particle's $\sigmasob$ along the column, 
it is lower than $\sigmagrid$ by a factor of $\sim 5$ in the high-density regions of interest. 
This suggests that, within each column, there are more particles with low $\sigmasob$;
the average, therefore, is biased towards a lower value.
Given that our current Sobolev approximation gives lower $\Sigma_{\rm{HI}}$ values, we could regard our result on star formation as a lower limit (See Section~\ref{sec:sf}). 
However, given that many $\sigmasob$ values are also found to be higher than the maximum $\sigmagrid$, it is possible that the factor of $\sim5$ difference between the two calculations is diminished:
the high $\sigmasob$ values may compensate for the mass-weighted averages being lower than $\sigmagrid$.
\citet{Kuhlen12} found qualitatively similar results when implementing a Sobolev-like approximation in their cosmological AMR simulations.  They were able to directly compare the Sobolev values to the true column density of their cells during runtime, and found the Sobolev approximation to return systematically lower column densities.  In the end however, they concluded that the differences were minor and did not have an overall impact on their final results.

We can then find the dust optical depth by the relation $\tau_c=\Sigma_{\rm{gas}}\sigma_d/\mu_{\rm{H}}$, where $\sigma_d$ is the dust cross section per hydrogen nucleus and $\mu_{\rm{H}}$ is the mean mass per H nucleus.  The dust cross section is taken to 
be $\sigma_{d,-21}=10^{-21}Z_{\rm{sn}}$ [cm$^2$], 
where $Z_{\rm{sn}}$ is the gas metallicity normalized to the solar neighborhood value $\Zsun=0.0204$ \citep{Rodriguez11}.  
The KMT model shows that, if the ISM is in a self-consistent two-phase equilibrium, then the ratio $G_0'/n_{\rm{H}}$ takes a characteristic value, and Equations~(\ref{eq:sval1}) \& (\ref{eq:chi1}) become
\begin{equation}
s\approx \frac{\ln\left(1+0.6\chi+0.01\chi^2\right)}{0.04 \, \left(\frac{Z}{\Zsun}\right)\left(\frac{\Sigma_{\rm{HI}}}{\msun \rm{pc}^{-2}}\right)}
\label{eq:sval}
\end{equation}
and 
\begin{equation}
\chi\approx3.1\frac{1+3.1\left(Z/\Zsun\right)^{0.365}}{4.1},
\label{eq:chi}
\end{equation}
respectively.  Using Equations~(\ref{eq:sval}) \& (\ref{eq:chi}) we can now calculate the $\htwo$ fraction of each gas particle by means of Equation~(\ref{eq:fh2calc}), if $0<s<2$ \citep{McKee10}, and otherwise $\fhtwo=0$.

\begin{figure}
\begin{center}
\includegraphics[scale=0.43]{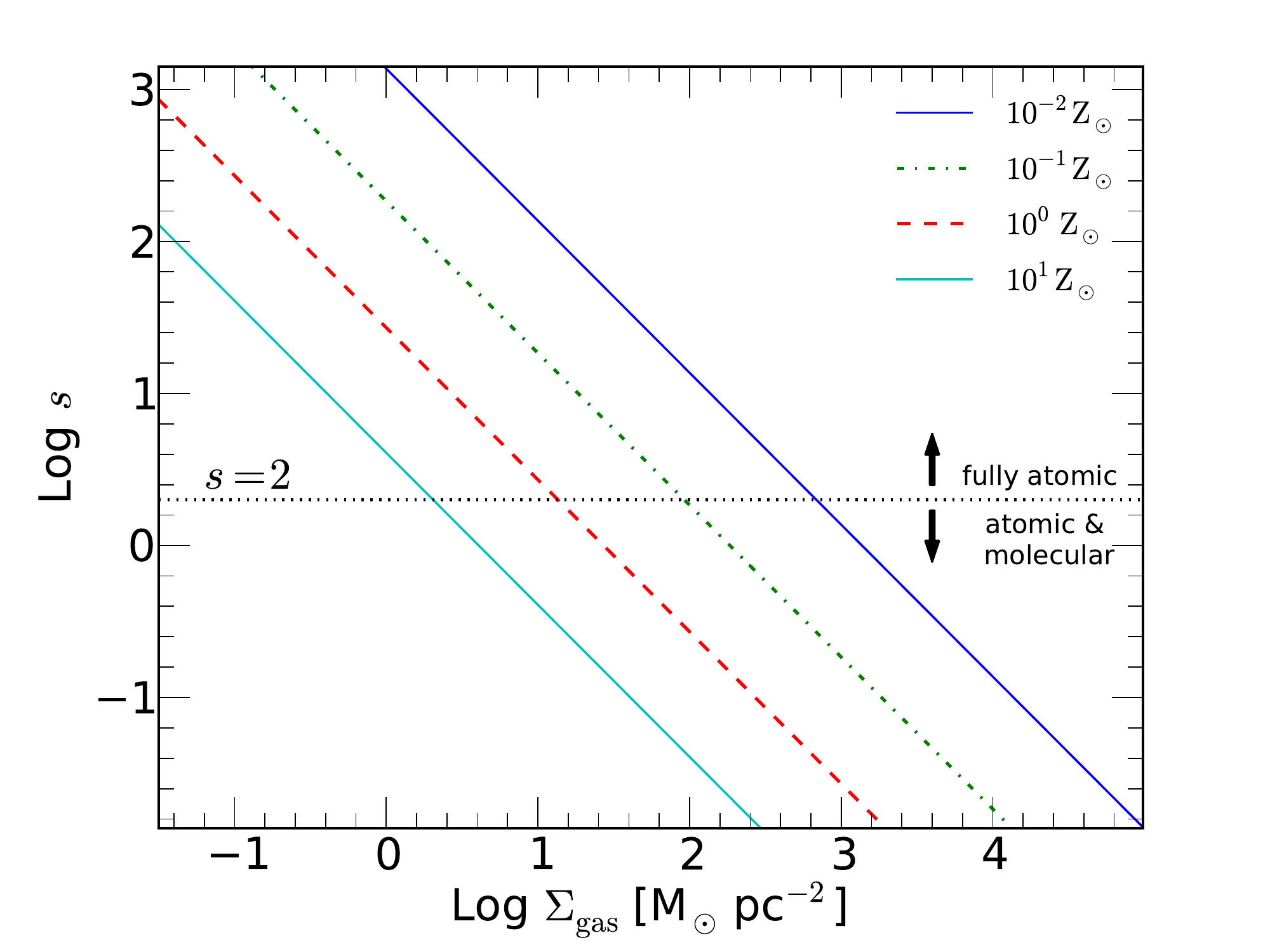}
\caption{
The parameter $s$ (Equation~\ref{eq:sval}) as a function of gas surface density for different metallicities.  
The value of $s=2$ corresponds to the transition from fully atomic gas to atomic \& molecular within the KMT model.
Lower metallicity gas requires larger column densities (i.e., more shielding) in order to form $\htwo$.
}
\label{fig:svalues}
\end{center}
\end{figure}

Figure~\ref{fig:svalues} shows how the transition from fully atomic to atomic \& molecular phase for different metallicities.  
The value $s=2$ is the minimum complex size required for the transition from atomic to molecular gas for any given metallicity.
This represents the need for the gas cloud to be sufficiently large to form $\htwo$.  
The external UV radiation is first absorbed by dust, which is essentially traced by the metallicity, then by a thin molecular region.
If the cloud is too small, then there will not be enough surrounding material to absorb all the UV photons, and the $\htwo$-core will be dissociated.  
If the cloud is large, we expect a large molecular core.
Higher metal content effectively absorbs more radiation, allowing for the formation of $\htwo$ at lower surface densities.

Knowing $\fhtwo$ allows us to modify our SF model by replacing $\rho_{cold}$ in Equation~(\ref{eq:springelsfr}) with the $\htwo$ mass density $\rho_{\rm H_2} = \fhtwo \, \rho_{\rm{HI}}$.
We also change our SF time-scale to a more physically realistic value, namely the free-fall-time of the $\htwo$ gas available to form stars
\begin{equation}
\tff = \sqrt{\frac{3\pi}{32 G \rho_{\rm{H}_2}}}.
\label{eq:tff}
\end{equation}

Here, it may be more correct to use $\rho$ as apposed to $\rho_{\rm{H}_2}$ in Equation~\ref{eq:tff}.  In the case where $\fhtwo\ll1$ for instance, we are representing a physical situation where most of the mass is in \HI~ and there is a small molecular cloud.  The molecular gas here will be at lower temperatures than the atomic gas with an internal volume density larger than what we would obtain via $\fhtwo\times\rho_{\rm{H}I}$.  In general, the internal volume density of the molecular cloud is greater than or equal to $\rho$, but to what degree depends highly on internal processes that we cannot model in our current simulations, such as the internal velocity dispersion of the cloud and the molecular cloud virial ratio (M. Krumholz, private communication).  In practice this choice does not matter very much anyway, because the bulk of star formation will occur in particles with $\fhtwo \simeq 1$ for either choice.
We have performed additional simulations to examine the difference between using
$\rho_{\rm{H}_2}$ or $\rho$ in the denominator of Equation~\ref{eq:tff},
and have confirmed that the differences are insignificant for the results presented in this paper.

Furthermore, observations have shown that SF is a slow process and that the efficiency at which dense regions produce stars is $\sim$1\% \citep{Krumholz07,Lada10}.  To account for this we introduce an efficiency parameter of $\epsilon_{\rm{SF}}=0.01$ which leads us to our new formulation of Equation~(\ref{eq:springelsfr}):
\begin{equation}
\dot{\rho}_*=(1-\beta) \, \epsilon_{\rm{SF}} \,  \frac{\rho_{\rm{H_2}}}{\tff}. 
\label{eq:h2sfr}
\end{equation}
If a gas particle has $\fhtwo > 0$, then it is eligible to form stars at the above rate. 

\subsubsection{Assumption on the neutral fraction}
\label{sec:neutralassumption}

As discussed in the previous section, in order to calculate the fractional $\htwo$ abundance, we must first calculate the scale-height of \HI, which then allows for the calculation of $\Sigma_{\rm{HI}}$.  Our \gadget code tracks the neutral fraction of each gas particle individually.  
For the high density multiphase gas however, the neutral fraction is tracked only for the hot phase, and the cold gas fraction $x \equiv \rho_c / \rho$ is computed within the multiphase ISM sub-particle model \citep{Springel03}. 
For the very high-density particles, most of the mass is in cold, neutral phase ($x \gtrsim 0.95$), but the tenuous hot phase determines the temperature.
We calculate the neutral fraction using the $x$-parameter for high-density particles above the SF threshold for our N144L10 fiducial run at $z=3$, and find all of particles to have $f_{\rm{HI}}>0.96$, and 98\% of the hydrogen mass to have $f_{\rm{HI}}>0.97$.  
Given the small mass fraction of ionized gas, it is a good approximation to assume that any gas particle with $n_{\rm th}^{\rm SF} > 0.6$\,cm$^{-3}$ \citep{Choi10} is completely neutral ($f_{\rm{HI}}=1$) for the scale-height and column density calculations detailed in Section~\ref{sec:H2model}.

\begin{figure}
\includegraphics*[scale=0.44]{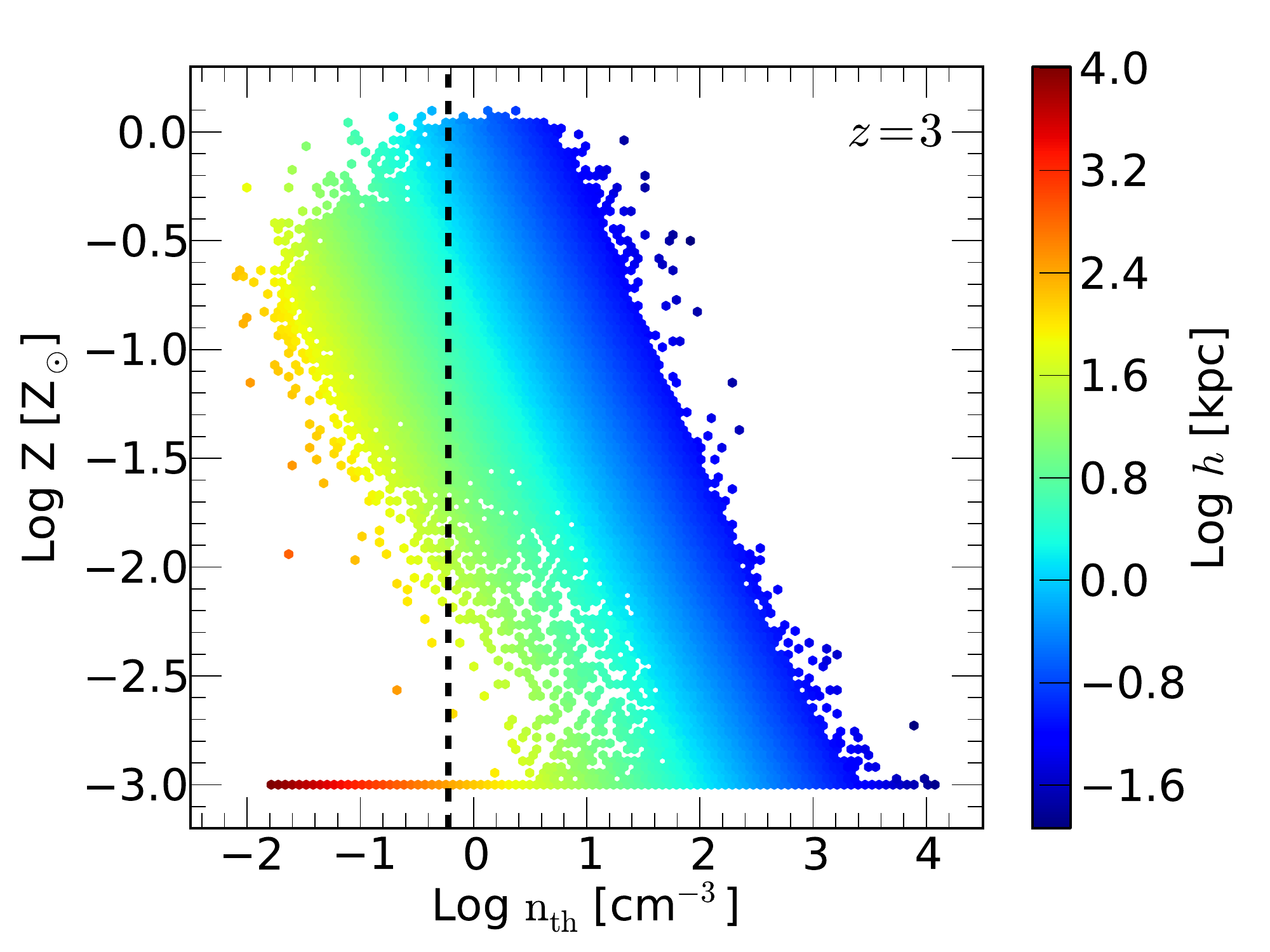}
\caption{
Physical density threshold (Eq.~[\ref{eq:adaptivethresh}]) for $\htwo$ formation of all particles within a low resolution run (N144L10) at $z=3$.   Color gradient corresponds to the median scale-height $h$ (Eq.~[\ref{eq:h}]) as indicated by the color bar.  The black dashed line represents the physical SF density threshold of $\rm{n_{th}^{SF}=0.6}$\,cm$^{-3}$ used in our previous SF models \citep{Choi10} .  
The KMT model generally requires higher densities to form $\htwo$ and hence be eligible for SF, compared to the Fiducial model. 
}
\label{fig:rhothresh}
\end{figure}

\subsubsection{$\htwo$ formation threshold density and Wind model modifications}
\label{sec:wind}

This new SF model (Eq.\,[\ref{eq:h2sfr}]) allows us to compute the density threshold ($\rho_{th}$) for individual particles based on their metallicity, above which results in the formation of $\htwo$.  
In the earlier version of our \gadget code, \citet{Choi11} implemented the ``Multi-component Variable Velocity" wind model,  
in which a particle was allowed to travel as a wind particle with no hydrodynamic forces applied 
as long as its physical density exceeded  
$\rm{n_{th}^{SF}}$.
The wind velocity of each particle was calculated based on the SFR of the galaxy that the particle belongs to. 
We can now revise this wind criteria to be based off of an individual particle's $\htwo$ formation threshold rather than a fixed value as in previous SF models.

The formation of $\htwo$ requires sufficient shielding, or else the molecule will be dissociated by UV radiation.  We can set the threshold for $\htwo$ formation for each particle by solving Equation~(\ref{eq:sval}) for $\Sigma_{\rm{gas}}$ using $s=2$;  
this value allows us to calculate the minimum $\Sigma$ for SF within our model:
\begin{equation}
\frac{\Sigma_{th}}{\msun \rm{pc}^{-2}}\approx \frac{\ln\left(1+0.6\chi+0.01\chi^2\right)}{2\times0.04 \left(Z/\Zsun\right)}.
\label{eq:sigmathresh}
\end{equation}

We can then convert this surface density threshold to a volume density threshold using Equation~(\ref{eq:sobolev}) for each particle, which leads us to the minimum volume density required to form H$_2$ at that particle's particular metallicity: 
\begin{equation}
\rho_{\rm{HI}_{th}} \approx \frac{\ln\left(1+0.6\chi+0.01\chi^2\right) \msun \rm{pc}^{-2}}{2\times0.04\left(Z/\Zsun\right) h}.
\label{eq:adaptivethresh}
\end{equation}
In other words, this is the $\htwo$ formation density threshold. 
In the present work, if the physical density of a gas particle is above its particular adaptive $\htwo$ formation threshold $\rho_{\rm{HI}_{th}}$, then it is eligible to be a member of the wind.
This minor modification is necessary as it allows the wind formation criteria to adapt to the new SF model.  
It is a natural consequence based on the assumption that the galactic outflows result from SF, but we will investigate its implication further in the future using higher resolution 
zoom or isolated galaxy simulations.

Figure~\ref{fig:rhothresh} shows the values of $\rho_{\rm{HI}_{th}}$ vs. metallicity of each particle
in a low resolution simulation (N144L10).
The previous SF density threshold is shown as the black dashed vertical line.  
Generally, the values of $\rho_{\rm{HI}_{th}}$ are higher for each particle than in the previous SF models, allowing for particles to reach higher densities before becoming eligible to form stars.
This makes SF in the $\htwo$ model less efficient than in the previous SF models. 
This plot also shows that for a given metallicity, a lower $h$ returns higher $\rho_{\rm{HI}_{th}}$.
Particles with higher metal content have lower formation thresholds due to their ability to absorb more dissociating photons.
The accumulation of particles at $Z = 10^{-3} \Zsun$ corresponds to those that have yet to be enriched by SF, but they have varying $\rho_{\rm{HI}_{th}}$ due to variations in $h$.  Some particles at $-2 < \log n_{\rm th} < 0$ have already been enriched by $z=3$. 

\begin{figure*}
\includegraphics*[width=1\textwidth,clip]{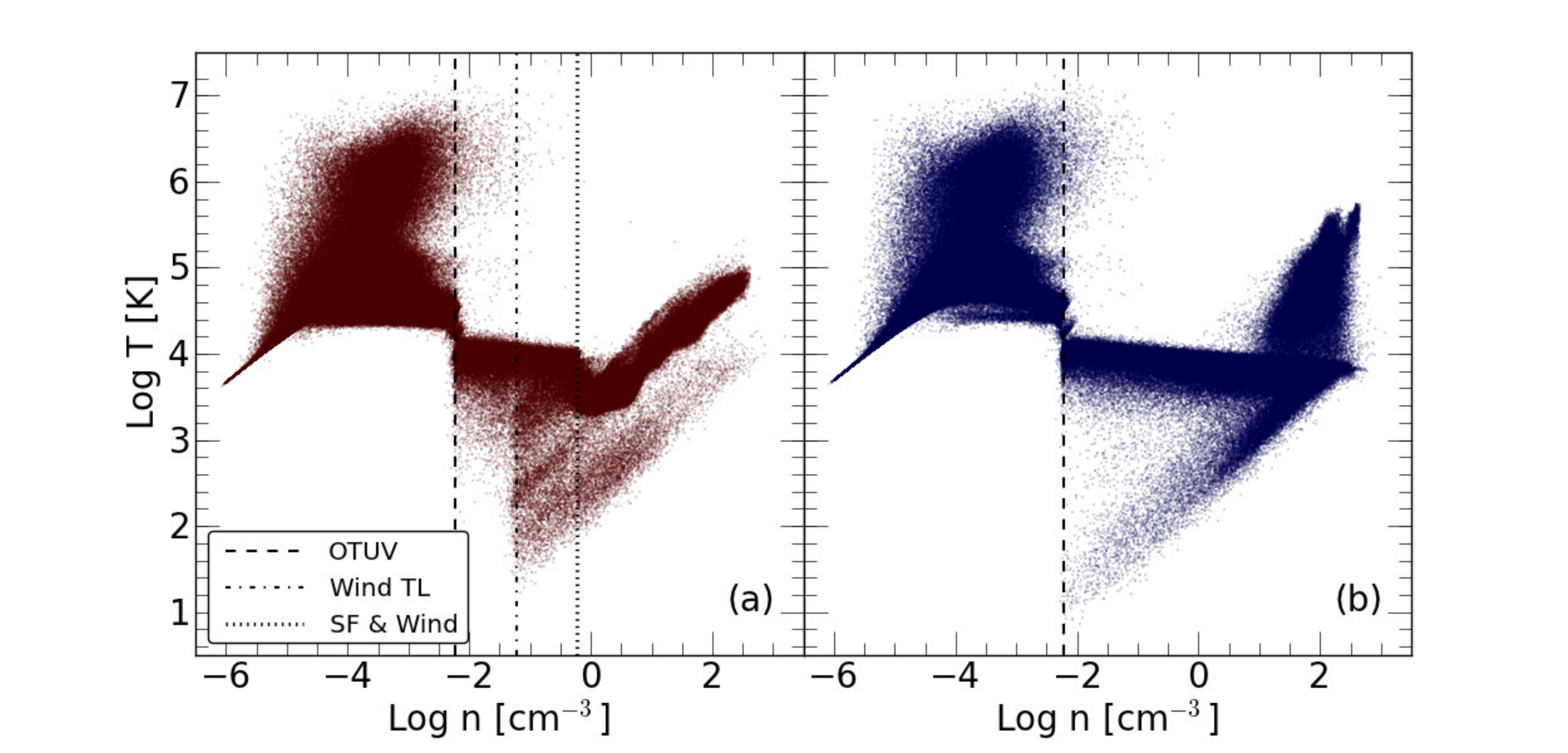}
\caption{Gas temperature vs. number density phase diagrams for the low resolution N144L10 runs at $z=3$.
{\it Panel (a)}:  the Fiducial run.  The right-most dotted line is the physical SF and wind density threshold ($\rm{n_{th}^{SF}}=0.6\rm{cm}^{-3}$, Section~\ref{sec:phase}),
the middle dot-dashed line is the maximum wind travel length (TL) discussed in Section~\ref{sec:phase},
and the left-most dashed line corresponds to the OTUV threshold discussed in Section~\ref{sec:phase}.
{\it Panel (b)}: $\htwo$ run.  Here only the OTUV threshold is shown.  There is no fixed SF density threshold, as it is different for every particle depending on the metallicity and surface density as described in Section~\ref{sec:wind}.
}
\label{fig:phasecompare}
\end{figure*}

\subsubsection{Metal Floor}
\label{sec:metalfloor}

In our Fiducial runs, initially all gas particles are metal free. 
Enrichment occurs during star formation; in this process SN explosions return a metal mass of $\Delta M_Z = y_* \Delta M_*$ to the ISM, where $y_*=0.02$ is the yield, and $M_*$ is the mass of newly formed long-lived stars.	
It is assumed that each gas particle behaves as a closed box locally, wherein metals are instantaneously mixed between cold clouds and ambient hot gas.

Within the framework of our new SF model, stars can only form if they contain $\htwo$, as determined by Equation~(\ref{eq:fh2calc}).  In order for $f_{H_2}\neq0$, the gas particle must have some metal content.
To begin forming stars, we must first enrich the gas particles by hand at very high redshift. 
Recent high resolution numerical studies by \citet{Wise12} found that a single pair-instability supernova of a Pop-III 
or very early Pop-II ($z>11$) star can enrich its host halo to a metallicity of $10^{-3} \Zsun$.
Their findings are in agreement with observed DLA metallicities, which have metal floors of the same order \citep{Wolfe05,Penprase10}.
To mimic this enrichment, we introduce a metal floor of $Z = 10^{-3} \Zsun$ for all gas particles at a specified epoch.

To test the impact of the assumed metal floor, 
a few low resolution simulations (N144L10) are performed introducing the metal floor at redshifts of $z=9,11$, \&13;  we refer to these as MF9, MF11, \& MF13 runs, respectively.
The cosmic SFRD histories are nearly identical between these three simulations;  
they differ only in the point at which each simulation starts to form stars.  This is due to the different times at which their metal floors are introduced.
The MF11 \& MF13 runs both start to form stars at $z\sim 9.2$, while MF9 does not begin star formation until $z\sim8.6$.
After their initial bursts of star formation, each of the three simulations begin following the same SFRD track from $z\sim8.3$ to $z=3$. 
The GSMF between the three simulations are also nearly identical at $z=3$ \& 6, suggesting that the formation of galaxies within our simulations does not heavily depend on when the metal floor is set.
Lastly, the SHMR (cf. Section~\ref{sec:SHMR}) is examined at $z=3$ \& 6 for each simulation.
We find significant scatter in the results for all three runs, but the median SHMR values for each simulation are all well within one standard deviation of one another.
This again suggests that the stellar fraction of each halo does not depend heavily on the time at which the metal floor is set.

The redshift at which the metal floor is introduced is related to the metal enrichment by Pop-III and very early Pop-II stars.  
Since the aim of the current simulations is not to study galaxy evolution at $z>11$,
we choose to introduce our metal floor at the epoch of reionization.
Observational estimates by \citet{WMAP7} suggest this happens at redshift $z=10.6\pm1.2$.
In our simulation, reionization is set by the UV background model \citep[][December 2011 version]{Faucher09}\footnote{\url{https://www.cfa.harvard.edu/~cgiguere/UVB.html}},  hence our metal floor of $Z = 10^{-3}\, \Zsun$ is set at $z\sim10.75$ accordingly. 


\subsection{Gas phase diagram}
\label{sec:phase}

We examine a low resolution N144L10 simulation to study the gas temperature-density phase diagram.  
Figure~\ref{fig:phasecompare} compares our Fiducial run with the new $\htwo$ run at $z=3$.
Panel~(a) represents our Fiducial run and contains three vertical lines representing different thresholds.  
The left most dashed line ($\rm{n_{th}^{UV}=0.006}$\,cm$^{-3}$) represents an optically-thick ultra-violet threshold or `OTUV' \citep{Nagamine10}.  Any particle below this threshold will be heated by the UVB to $>10^4$\,K; any particle above this threshold is shielded from the UVB.
\citet{Nagamine10} and \citet{Yajima11} have demonstrated that this threshold is reasonable using full radiative transfer calculations. 

The right most dotted line in Figure\,\ref{fig:phasecompare}a represents the constant SF physical density threshold ($n_{\rm th}^{\rm SF}=0.6$\,cm$^{-3}$) in the Fiducial run. 
Any particle whose density exceeds this threshold is allowed to form stars based on the prescription outlined in Section~\ref{sec:pressuremodel}.
At densities \& temperatures above $n \sim 3$\,cm$^{-3}$ \& $T\sim10^4$\,K, we begin to see the effects of SN feedback and the multi-phase ISM model \citep{Springel03}.
The cold phase component dominates the mass of the particle, but the hot phase governs the temperature.  What is seen in this `arm' is the temperature of the gas heated by SN feedback (hot phase component), and the density of the cold phase component.  When we direct our attention to the $\htwo$ run (Panel~b) we see that this arm is now extended out to higher densities at lower temperatures.  This is because stars are only allowed to form if the gas particles contain any $\htwo$ above the adaptive density threshold $\rho_{\rm HI,th}$ given by Equation~(\ref{eq:adaptivethresh}). 
As discussed earlier, Figure~\ref{fig:rhothresh} illustrates how $\rho_{\rm HI,th}$ is typically higher than the previous SF density threshold, which allows particles to condense to higher densities without being heated by SN feedback.

The dot-dashed line in between the two previously discussed lines in Figure~\ref{fig:phasecompare}a represents the maximum wind travel length (TL) threshold of $n_{\rm TL}^{\rm w} = 0.1 \times n_{\rm th}^{\rm SF} = 0.06$\,cm$^{-3}$. 
If a particle becomes a member of the wind, hydrodynamic forces are turned off for a brief period of time \citep{Springel03,Choi11}.
This allows the gas to adiabatically expand and cool to lower temperatures until the density drops below $n_{\rm TL}^{\rm w}$, or the brief period of time has elapsed.
The dot-dashed line is absent from the Panel (b) because of the varying density threshold for each particle.
Instead of a temperature discontinuity, as can be seen in Panel (a), we have a `tail' that extends all the way to the OTUV threshold.  This tail corresponds to wind particles that have adiabatically expanded as part of the galactic wind.  

\begin{figure}
\includegraphics*[scale=0.46]{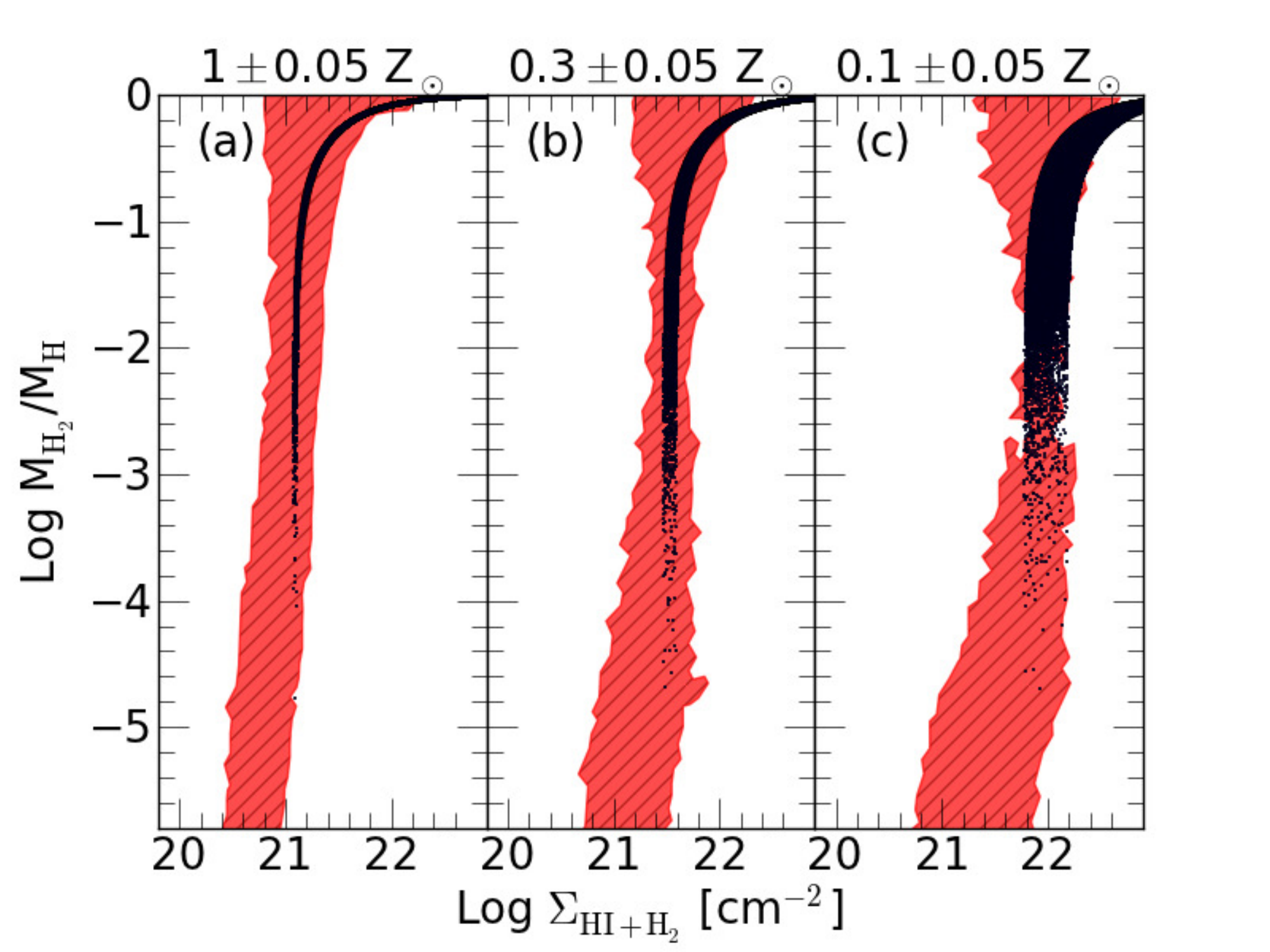}
\caption{
$\htwo$ mass fraction as a function of \HI+$\htwo$ surface density within our N600L10 run (black points) at $z=0$ for three different metallicity ranges.
The red shaded regions show the results of three Milky Way-like simulations of \citet{Christensen12} using a full non-equilibrium $\htwo$ model with different metallicities of (a) $1\Zsun$, (b) $0.3\Zsun$, and (c) $0.1\Zsun$.
}
\label{fig:sigmaH2}
\end{figure}

\subsection{Atomic to molecular transition}

It is important to study where the atomic to molecular transition occurs in our simulations.
Figure~\ref{fig:sigmaH2} shows this transition as a function of gas surface density in our N600L100 run at $z=0$ for three different metallicity ranges.
In the KMT model, the value of $\fhtwo$ is solely determined by the surface density of gas and metallicity (Equations~\ref{eq:fh2calc}, \ref{eq:sval} \& \ref{eq:chi}), therefore the simulation data (black dots) in all three panels is restricted to a relatively thin band determined by the range of metallicity chosen.

\citet{Christensen12} examined this transition for three isolated Milky Way-like simulations at different metallicities to test their fully time-dependent, non-equilibrium $\htwo$ calculation. Their raw simulation output can be seen as the red contour in Figure~\ref{fig:sigmaH2}.
The transition in our simulations is in good agreement with theirs, corroborating the comparison work of \citet{KrumholzGnedin11}.  
However, our data deviates to higher column densities at high molecular fractions due to the per-particle overestimation of the column density calculated by the Sobolev-like approximation, as discussed in Section~\ref{sec:H2regsf}.
This deviation should not pose a problem since particles in these high column density regions are already primarily molecular.

\section{Results on galaxy populations}
\label{sec:simresults}

\subsection{Dark matter halo content}
Dark matter (DM) particles were grouped using a simplified version of the parallel friends-of-friends group finder {\small SUBFIND} \citep{Springel01}.  The code groups the particles into DM halos if they lie within a specified linking length. 
We adopt a standard value of $b=0.2$ for the linking length parameter, which is a fraction of the initial mean inter-particle separation.  Each halo must also have a minimum of $100$ particles to be considered a halo.

The DM halo mass functions were found to be in agreement between the $\htwo$ and Fiducial runs. 
This is an expected result, because both sets of simulations were started from identical ICs. 
Both results are in good agreement with the \citet{ST99} mass function. 
This paper focuses on baryonic properties, and it is useful to examine and compare the contents 
of these halos between the Fiducial and $\htwo$ runs.
The contents of each halo are calculated by the summation of particle properties located within $r_{200}=[(G\mdm)/(100\,\Om(z)H(z)^2)]^{1/3}$ \citep{MoWhite02} of each halo's center of mass.

\begin{figure}
\includegraphics*[scale=0.37]{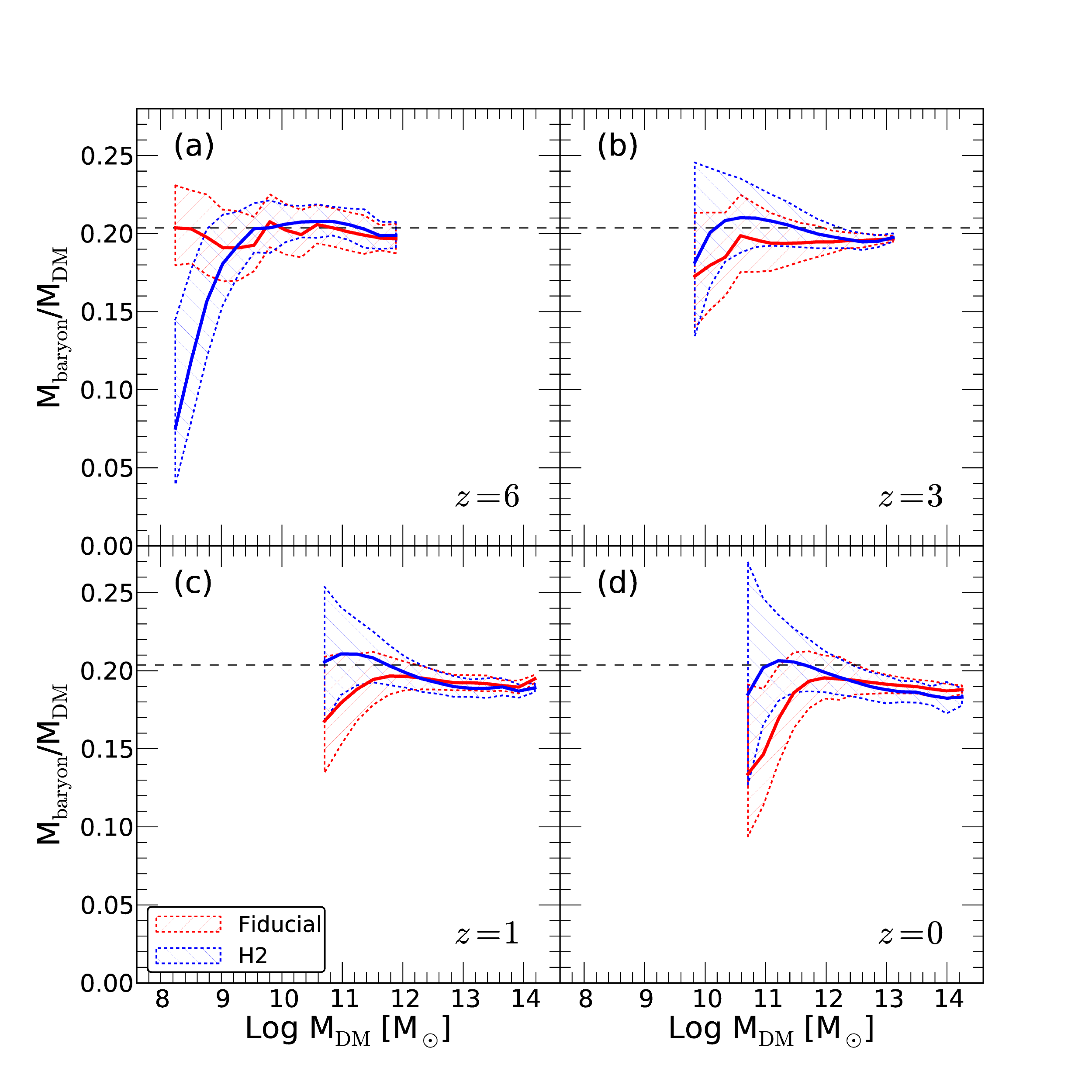}
\caption{
Baryon mass fraction within $r_{200}$ of each halo, ${\fb \equiv M_{\rm baryon} / \mdm = (M_{\rm gas,200} + M_{\star,200}}$)/$\mdm$, as a function of halo mass (only DM) for $z=6,3,1$, \& 0.  
The red and blue solid lines represent the median points in each mass bin for the Fiducial and $\htwo$ runs, respectively.  
The hatched regions represent  1$\sigma$ scatter in each $\mdm$ bin. 
The cosmic mean baryonic fraction ($\Omega_{\rm{b}}/\Omega_{\rm{DM}}$) is represented by the dashed horizontal line.
}
\label{fig:mhalombaryon}
\end{figure}

\subsubsection{Baryon fraction}
\label{sec:baryonicfraction}

Figure~\ref{fig:mhalombaryon} presents the baryon mass fraction over halo mass ($\fb \equiv M_{\rm baryon} / \mdm = (M_{\rm gas,200} + M_{\star,200}$)/$\mdm$) as a function of $\log \mdm$ at $z=6,3,1$ \& 0.
Here the cosmic mean ($\Omega_{\rm{b}}/\Omega_{\rm{DM}}$) is illustrated by the horizontal black dashed line.
Panel~(a) shows the composite data from the N400L10, N400L34, \& N600L100 runs at $z=6$; Panel~(b) is composite data from the N400L34 \& N600L100 runs at $z=3$; Panels~(c) \& (d) are composed of data from the N600L100 simulation only at $z=1$ and 0, respectively.
Solid lines represent the median value within each bin, and the hatched regions represent a 1$\sigma$ spread in the data. 
The cutoff of the data at lower mass end is determined by the halo mass limit of the halo grouping. 

At $z=6$ (panel~[a]), the $\fb$ of the two SF models agree with each other well and with the cosmic mean for halos above $\mdm \sim 10^{9} \msun$. 
Halos below this mass in the $\htwo$ run have lower $\fb$ than in the Fiducial run by $\sim35$\%.
This is presumably due to the different UVB model between the two runs; in the $\htwo$ run the UVB is turned on much earlier, and the gas in low-mass halos are photo-evaporated. 

This large difference in $\fb$ is nonexistent in low mass halos at $z=3$ as shown in Panel~(b).
The median values within the $\htwo$ run are now generally higher than those in the Fiducial run. 
As we will see in later sections, this is likely due to higher SFRs and hence stronger SN feedback 
in the Fiducial run, and this trend continues to $z=0$. 
The scatter in $\fb$ at $\mdm \sim 10^{9.7}\msun$ is greater for the $\htwo$ model, but it encompasses that of Fiducial run.  
Both begin to drop slightly below the cosmic mean at around $\mdm \sim10^{12}\msun$.

By $z=1$, $\fb$ in the most massive halos settle to a value that is lower than the cosmic mean by $\sim 7$\%. 
Again in Panel~(c), we see $\fb$ in massive halos with $\mdm > 10^{12}\msun$ is in agreement between the two models.  
At lower $\mdm$, the Fiducial run still shows a lower baryon fraction.
Finally at $z=0$, we see that, for the halos with $\mdm < 10^{12}\msun$, the mean $\fb$ has decreased slightly since $z=1$. 
This means that the halos on average have acquired more dark matter than baryons (either through mergers or accretion), and/or at the same time lose the gas from galaxies due to SN feedback. 
The scatter of $\fb$ for lower mass halos is generally greater than for the massive halos, and this is a known resolution effect \citep[e.g.,][]{Oshea05}.  

\begin{figure*}
\begin{center}
\includegraphics*[scale=0.5]{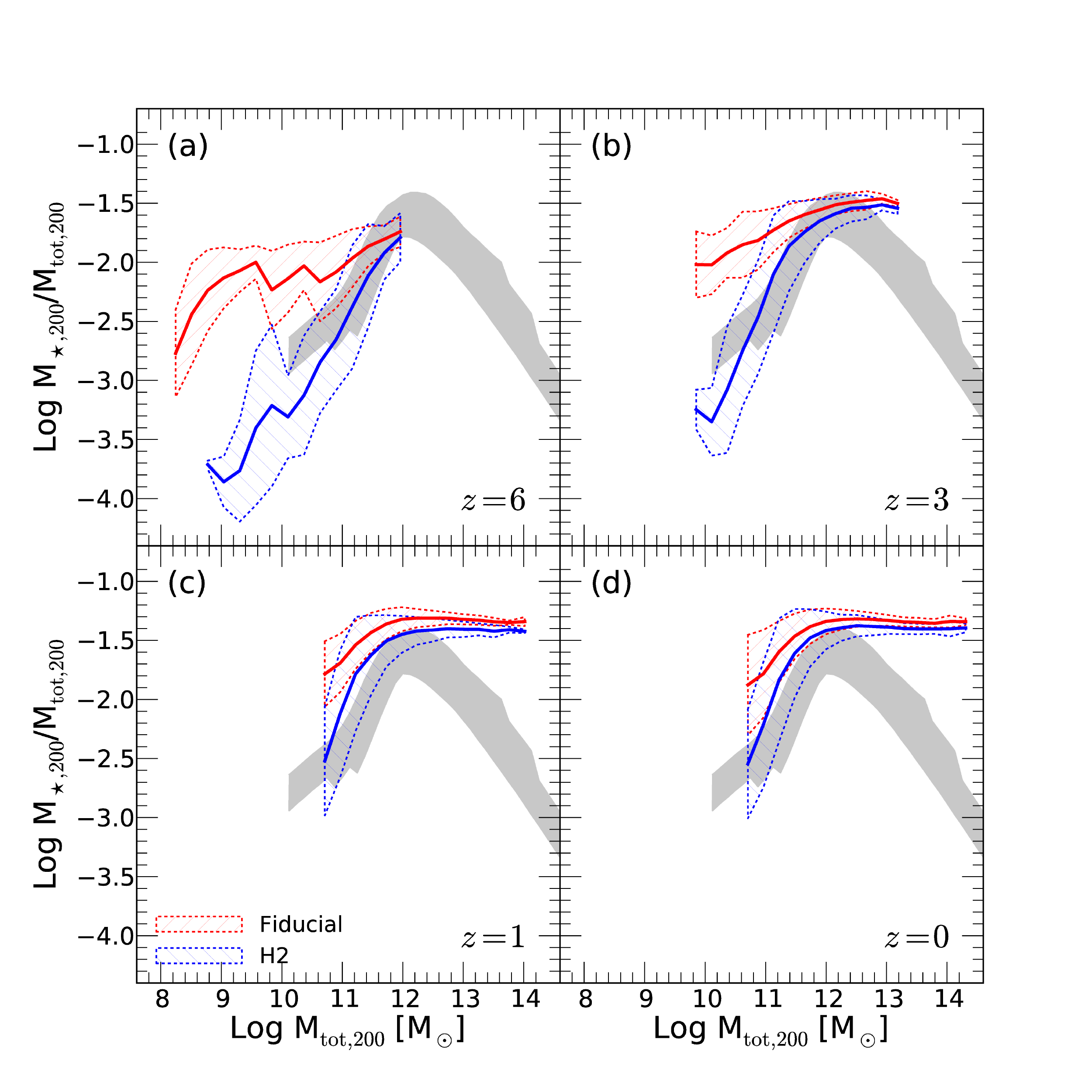}
\caption{
The SHMR as a function of total halo mass (DM+baryons) within $r_{200}$. 
The data from semi-analytic models and observations are shown as the grey shade, which is identical in all four panels as it doesn't evolve very much with redshift \citep{Behroozi13}.
}
\label{fig:SMHM}
\end{center}
\end{figure*}

\subsubsection{Stellar-to-halo mass ratio (SHMR)}
\label{sec:SHMR}

\begin{figure*}
\includegraphics*[scale=0.37]{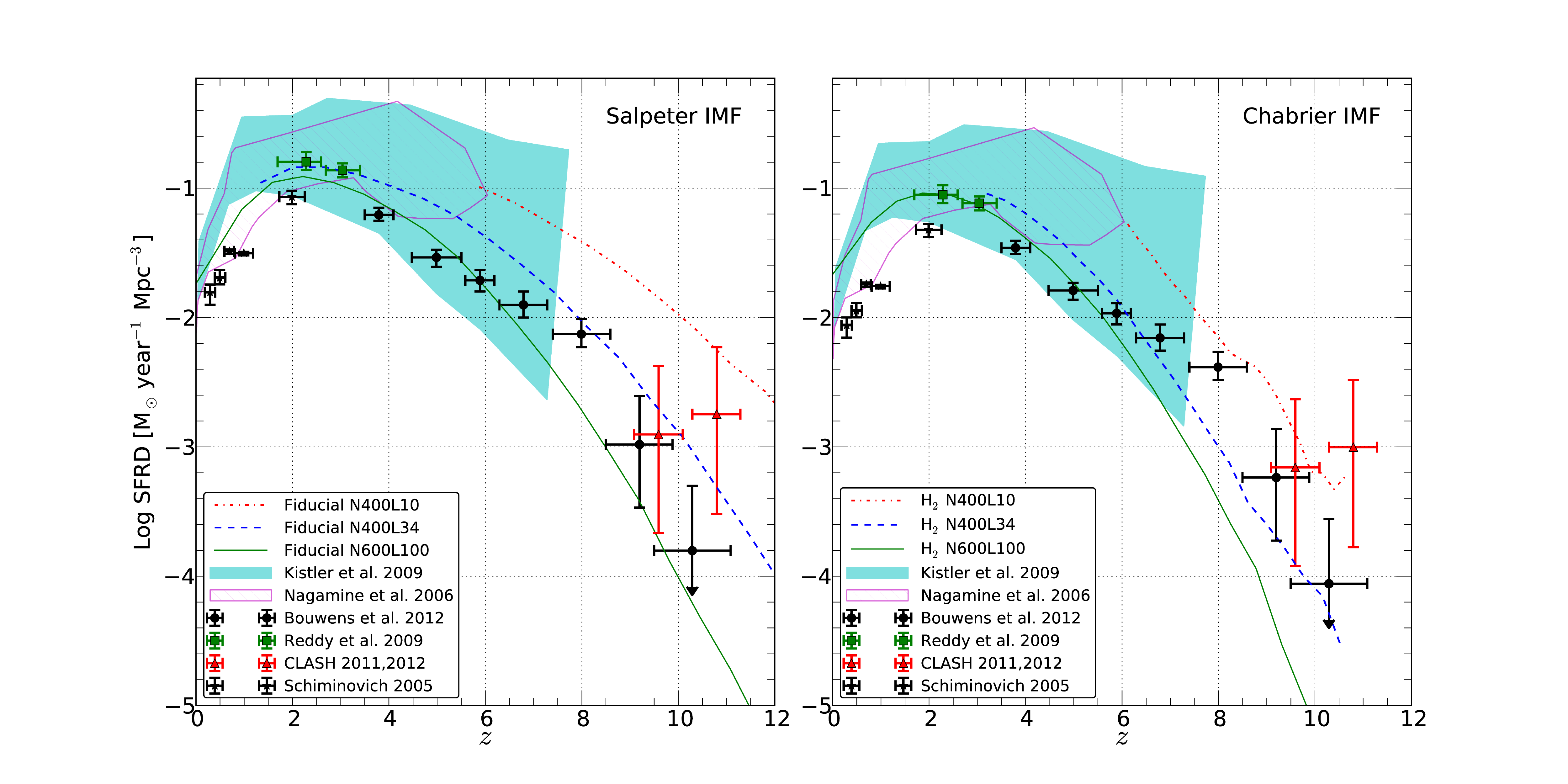}
\caption{Cosmic SFRD for our simulations compared to some observations. 
The left panel is for the Salpeter IMF (Fiducial runs), and the right for Chabrier IMF ($\htwo$ runs). 
The observational data are from: the CLASH program \citep[][red triangles]{Postman12,Coe12},
\citet[][black circles]{Bouwens11,Bouwens12}, \citet[][green crosses]{Reddy09}, \citet[][black stars]{Schiminovich05}, \citet[][cyan shade]{Kistler09}, and \citet[][magenta hatched region]{Nagamine06}.
All observational data are corrected for dust extinction by each author as they deemed appropriate. 
}
\label{fig:multimadau}
\end{figure*}

The ratio of stellar-to-halo mass as a function of total halo mass $\mtot$ (DM+gas+stars within $r_{200}$) provides a useful insight on the efficiency of star formation in different halos, and it has collected significant attention in the recent years \citep{Conroy07, Baldry08, Behroozi10, Moster10, Foucaud10, Evoli11, Leauthaud12, Papastergis12}. 
All of these work find a prominent peak in this relation at $\mtot \sim10^{12}\msun$, suggesting that there is a characteristic halo mass that galaxy formation is most efficient.  This mass-scale roughly corresponds to the characteristic stellar mass $M^*$ of GSMF, i.e., the knee of Schechter function, where most of the stellar mass has formed globally. 
To further surprise, \citet[][hereafter B12]{Behroozi13} found that, using observational data and semi-analytic modeling, SHMR is almost time-independent between $z=4$ to $z=0$.  This is interesting, because SHMR should reflect all the cumulative effects of past star formation and feedback, and a non-evolving SHMR suggests tight self-regulation and a subtle balance between star formation and feedback.
It is a significant challenge for any cosmological hydrodynamic simulation to reproduce this relation across a wide range of halo mass and cosmic time. 


Note that we are specifically using $\mtot$, and not $\mdm$ for this comparison.
This is because  \citet{Munshi12} pointed out that, in order to accurately compare simulations to semi-analytic model results (such as abundance matching technique), one must observe the simulations in a similar manner.
They refer to the work by \citet{Sawala12}, who showed that $\mtot$ in $N$-body simulations can be greater than those in $N$-body+hydro simulations by up to $30\%$, because various baryonic processes 
(gas pressure, reionisation, supernova feedback, stripping, and truncated accretion) 
can remove baryons from the halo, decreasing the total mass systematically. 
Additionally, \citet{Graham05} found that the stellar mass component of observed galaxies could be systematically underestimated by $\sim20\%$ in some cases; for example, additional flux of 0.22 mag lies beyond the SDSS Petrosian aperture for a galaxy that has an $R^{1/4}$ profile.  
Here, we consider that it would be more natural to examine SHMR as a function of $\mtot$ rather than correcting our results by a certain number. 

In Figure~\ref{fig:SMHM}, we show the SHMR in our simulations by calculating the total stellar mass contained within $r_{200}$ of each halo's center-of-mass ($M_{\star,200}$).
If we assume that the B12 data extends out to $z=6$, we see in Figure~\ref{fig:SMHM}a that our $\htwo$ run does a good job at reproducing the B12 data at $\mtot < 10^{12} \msun$, much better than the Fiducial model. 
We have verified that the different UVB models do not impact this result.
The comparison of the models in Figure~\ref{fig:SMHM} clearly suggests that the suppression of star formation in low-mass halos is favorably achieved by the $\htwo$ model.  Note that this SF suppression is not due to the SN feedback, but rather due to the metallicity dependence of the new $\htwo$-based SF model. 
This could become a critical point to distinguish between the $\htwo$-based and turbulence-based SF models in the future.

At $z=3$ (Figure~\ref{fig:SMHM}b) the SHMR increases slightly for both simulations.  
Our simulations are in agreement with the B12 data at $\mtot < 10^{12.2}\msun$.
However, we continue to overestimate SHMR at $\mtot > 10^{12.2}\msun$ down to $z=0$, which could be due to lack of AGN feedback in our current simulations.  It is widely believed that both thermal and momentum feedback from supermassive black holes suppresses the star formation in massive halos, making them `red \& dead' \citep[e.g.,][]{DiMatteo05, Springel05b, Nagamine05, Ostriker10, ChoiE12}. 
There is little evolution between $z=1$ \& 0 in our simulations (Figure~\ref{fig:SMHM}c,d), and 
our results are higher than B12 data even for low-mass halos at $z\le 1$.

\subsection{Quantities related to star formation}
\label{sec:sf}

\subsubsection{Cosmic star formation history}
\label{sec:sfrd} 

With the $\htwo$ model producing less stars in lower mass halos,  we expect the cosmic SFRD to be lower as well.
When comparing simulations to observational estimates of SFRD, we have to be careful about which IMF is being used. 
The Fiducial and $\htwo$ runs use different IMFs.  
In order to fairly compare the two, we must make corrections to either the simulation data or the observations.  Because SFR is a raw output of our simulation, we prefer to take the latter route and correct the observational data to the same IMF as in simulation. 
A simple factor $f_{\rm IMF}$ allows for this conversion:
\begin{equation}
\dot{\rho}^{\rm IMF}_\star=\dot{\rho}^{\rm Salpeter}_\star/f_{\rm IMF},
\label{eq:imfconvert}
\end{equation}
where $\dot{\rho}_\star^{\rm IMF}$ represent an arbitrary IMF.
To convert from Salpeter to Chabrier, we take $f_{\rm IMF}=1.6$ \citep[e.g.,][]{Nagamine06, Raue12}, and from Salpeter to Kroupa we take $f_{\rm IMF}=1.8$ \citep{Horiuchi09}.
This is because, for a given amount of observed rest-frame UV flux, the required SFR would be lower for an IMF that is weighted more towards higher masses. 
After correcting our IMFs, we find that both simulations roughly agree with the observed data.

Figure~\ref{fig:multimadau} shows the cosmic SFRD history as a function of redshift. 
As expected, the $\htwo$ runs show significantly lower SFRD at most redshifts relative to the corresponding Fiducial runs.
The $\htwo$ runs do not start forming stars until $z\sim10.5$, which is a consequence of our model.
In the $\htwo$ run, in order for gas to be eligible for SF, it must first contain $\htwo$, which requires non-zero metal content. 
As discussed in Section~\ref{sec:metalfloor}, we introduce the metal floor by hand at $z\sim 10.75$, after which stars are able to form.  Until the metal floor is introduced, the gas continues to condense to higher densities. 

The slopes of the $\htwo$ SFRDs at high-$z$ are slightly steeper than the Fiducial runs, because the $\htwo$ run 
starts SF a little later than the Fiducial run, and has more abundant high-density gas available for SF.  
It tries to catch up to the Fiducial run after the metal floor is introduced. 
For the same reason, the peak of the SFRD of the N600L100 $\htwo$ run is slightly shifted towards lower redshift compared to the Fiducial run.  With a lower SFRD in the $\htwo$ runs, there is more gas available for SF at later times.

Note that Figure~\ref{fig:multimadau} is only showing the results of different simulations as separate lines, and it does not show the expected total SFRD.  To really obtain the total SFRD, we must carefully examine the contribution from each galaxy mass ranges to SFRD, and sum up the contribution from each simulation.  This was done by \citet{Choi10} for the Fiducial runs, and we will present such analyses separately \citep{Jaacks13}.
Here, we simply note that the expected total SFRD would be even higher than the red dot-dashed line of N400L10 run, and it would go right through the data range shown by the cyan and magenta shaded regions. 


\begin{figure}
\includegraphics*[scale=0.36]{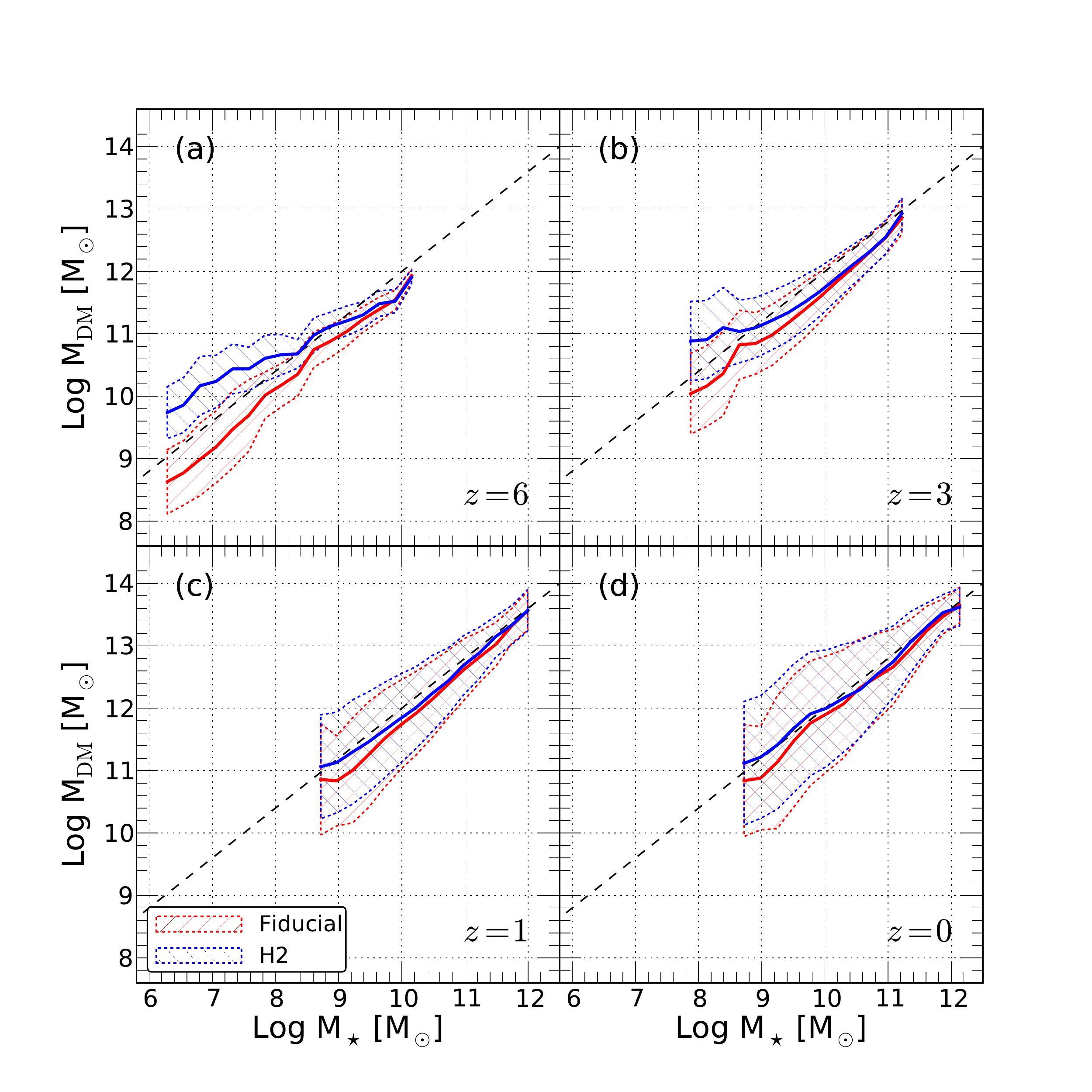}
\caption{Relationship between the masses of simulated galaxies (identified by the friends-of-friends grouping) and their nearest DM halos.  
Note that $\mstar$ is not exactly same as $M_{\star, 200}$. 
We see that the low-mass galaxies with $\mstar \sim10^6 - 10^9\msun$at $z=6$ reside in more massive DM halos in the $\htwo$ runs than in the Fiducial run.  The dashed line in each panel represent the scaling of $\log \mdm = 0.8 (\log \mstar -10) + 12$, which is a simple eye-ball fit at $z=0$. 
}
\label{fig:mstarVSmhalo}
\end{figure}

\begin{figure}
\includegraphics*[scale=0.36]{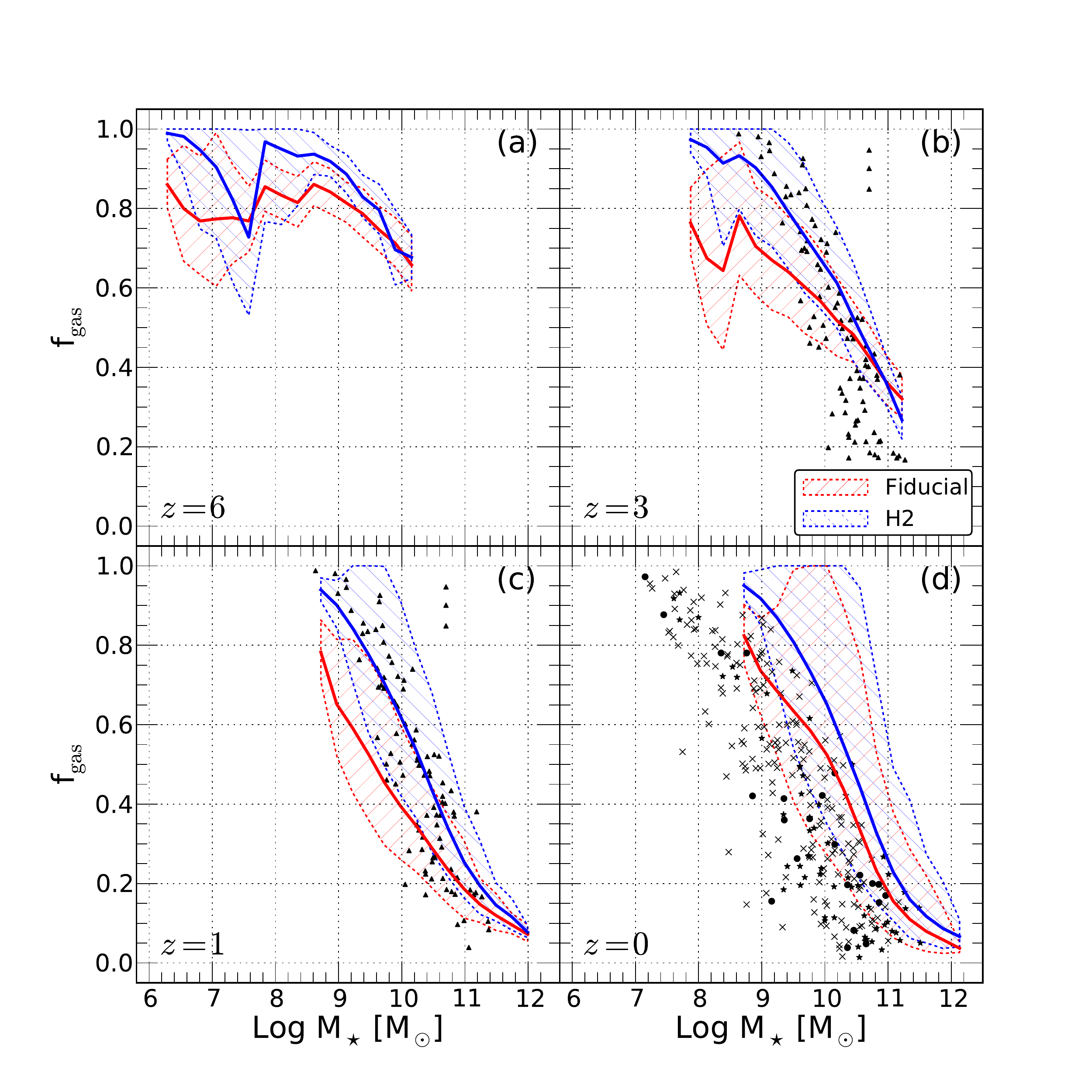}
\caption{
Gas mass fraction $\fgas=\mgas / (\mgas+\mstar)$ of simulated galaxies as a function of galaxy stellar mass.
Black triangles in Panels~(b)~\&~(c) are observed galaxies at $z\sim2$ \citep{Erb06}.  In Panel~(d) we show observational data at $z=0$ taken from \citet{Peeples11}.
}
\label{fig:mstarfgas}
\end{figure}

\subsubsection{In which halos do galaxies sit?}

So far, we have not considered the grouping of galaxies themselves (i.e., star and gas particles).
To examine individual galaxies in our simulations, we group gas and star particles based on the baryonic density field rather than the dark matter. 
This allows us to identify galaxies directly, and then calculate properties such as their SFRs, stellar masses ($\mstar$, which we distinguish from $M_{\star,200}$), gas masses ($\mgas$, which we distinguish from $M_{\rm gas,200}$)
and metallicities.  A more detailed description of this galaxy group finding process can be found in \citet{Nagamine04groupfinder}.  

Together with the friends-of-friends halo finding, we are interested in how the grouped galaxies relate to the DM halos. 
To find out the matching between the two sets, we search for the nearest DM halo from the center-of-mass of each galaxy. 
We limit our galaxy sample to those with minimum 10 star particles, and those that reside in halos with at least 100 DM particles.
Note that the DM structure between the Fiducial and $\htwo$ runs are nearly identical, because they both use identical ICs.

We can then make a scatter plot of corresponding $\mdm$ and $\mstar$ of each halo as shown in Figure~\ref{fig:mstarVSmhalo}.
In Panel~(a) we see that the low mass galaxies ($\mstar \sim10^6 - 10^9\msun$) at $z=6$ in the two runs reside in different halos with different masses; the median result of the two runs lie almost an order of magnitude apart, with the Fiducial run galaxies residing in lower mass halos on average. 
This is because the SF requires a higher threshold density in the $\htwo$ run and the gas requires a deeper potential-well of massive halos in order to form the same amount of stars as in the Fiducial run. 
The results of the two runs converge at higher masses ($\mdm >10^{9}\msun$), suggesting that those halos contain similar galaxies in the two runs.  For higher mass halos with $\mdm > 10^{11.5}\msun$, there seems to be little difference in the SF between the two runs. 

This $\mstar - \mdm$ relation does not evolve significantly as time proceeds. 
For comparison, the dashed lines in Figure~\ref{fig:mstarVSmhalo} represent the same scaling of  $\log \mdm = 0.8 (\log \mstar -10) + 12$.  The figure shows that the halos grow in mass with time, and the median lines slide up to top-right direction along the dashed line.

\subsubsection{Gas mass fraction of simulated galaxies}
\label{sec:gasfraction}

\begin{figure*}
\begin{center}
\includegraphics*[scale=0.5]{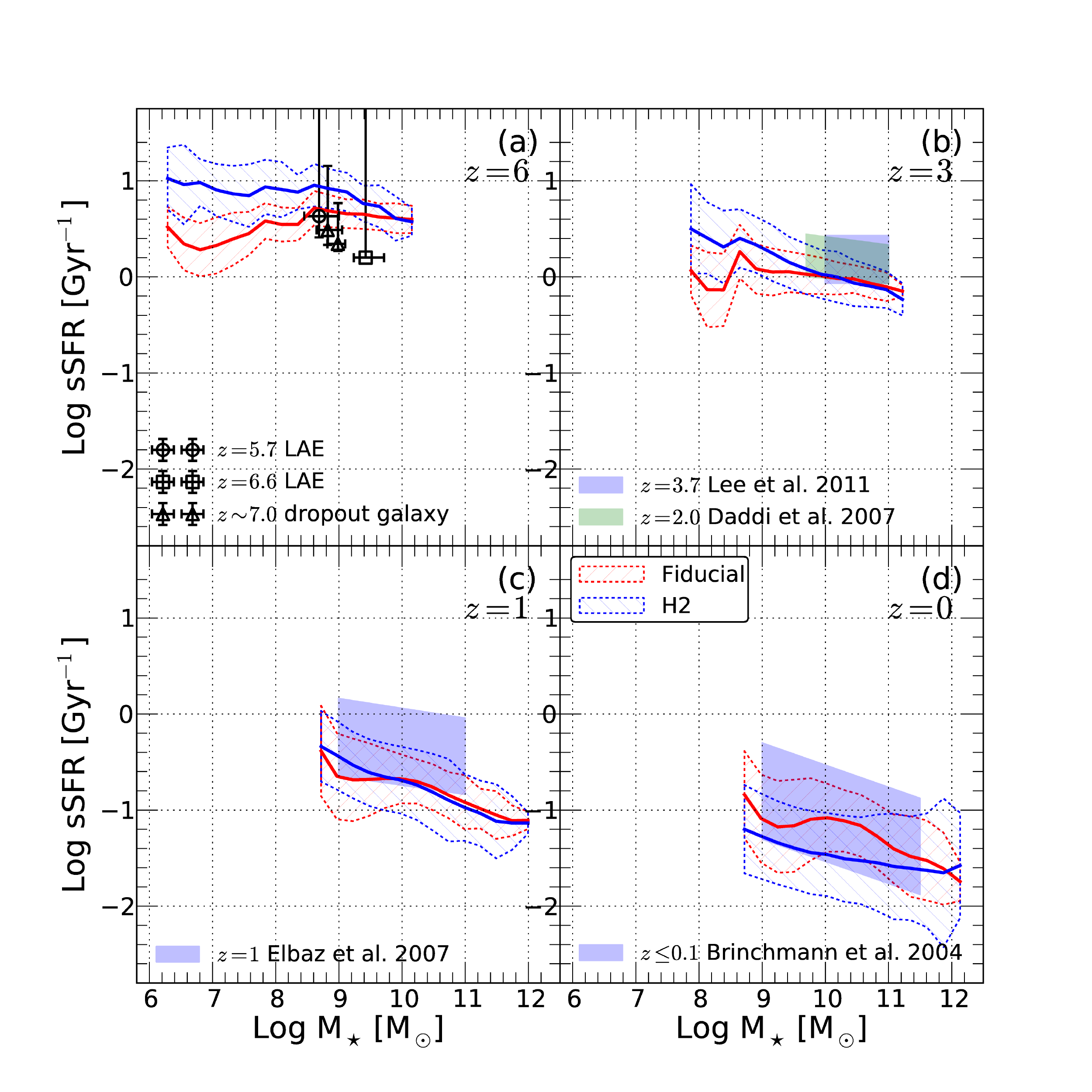}
\caption{
Specific star formation rate (${\rm sSFR \equiv SFR / \mstar}$) of simulated galaxies as a function of galaxy stellar mass.  The observed data ranges are indicated by the shaded region with the source indicated in each panel. 
Contoured observational data was taken from \citet{Krumholz11}, while the observations at $z=6$ were taken from \citet[][circles]{Ono10} and \citet[][triangles]{Labbe10}.
}
\label{fig:mstarsSFR}
\end{center}
\end{figure*}

Figure~\ref{fig:mstarfgas} shows the median gas fraction ($\fgas \equiv \mgas/(\mgas+\mstar)$) of simulated galaxies as a function of galaxy stellar mass.
In general, $\fgas$ in the $\htwo$ run is higher than that in the Fiducial run, but the 1$\sigma$ regions overlap with each other. 
The non-smoothness in the median lines in Panels (a) \& (b) are simply due to the procedure of combining data from 
multiple simulations of different resolutions. 

We find that the median lines do not evolve very much over time.  
At $z\le 3$, $\fgas$ declines steeply with $\mstar$; from values close to unity at $\mstar \sim 10^9\msun$ to $\fgas \lesssim 0.1$ at $\mstar \sim 10^{12}\msun$. 
This suggests that the massive galaxies with $\mstar \sim 10^{12}\msun$ in our simulations have converted most of their baryons into stars, and not much gas is left in them, coinciding with the downturn of the SFRD at these epochs (Figure~\ref{fig:multimadau}).

Black triangles in Panels~(b)~\&~(c) are from a sample of galaxies at $z\sim2$ \citep{Erb06}.
Simulated galaxies from the $\htwo$ run tend to agree better with the observed data at $z=3\,\&\,1$.
In Panel~(d) we show observational data of nearby galaxies from \citet[][stars]{McGaugh05}, \citet[][filled circles]{Leroy08}, and \citet[][crosses]{West09,West10}.  Neither the $\htwo$ or Fiducial models agree well with observations at $z=0$.  This may in part be due to the limited mass resolution of the N600L100 run; a higher resolution run would resolve lower mass galaxies, possibly shifting the distribution to the left in better agreement with observations, if we were to make a composite plot from different runs. 
Another possible cause for this discrepancy is that too much unenriched gas has fallen into these massive galaxies between $z=1$ and $z=0$, pushing $\fgas$ to higher values.

\begin{figure}
\includegraphics*[scale=0.43]{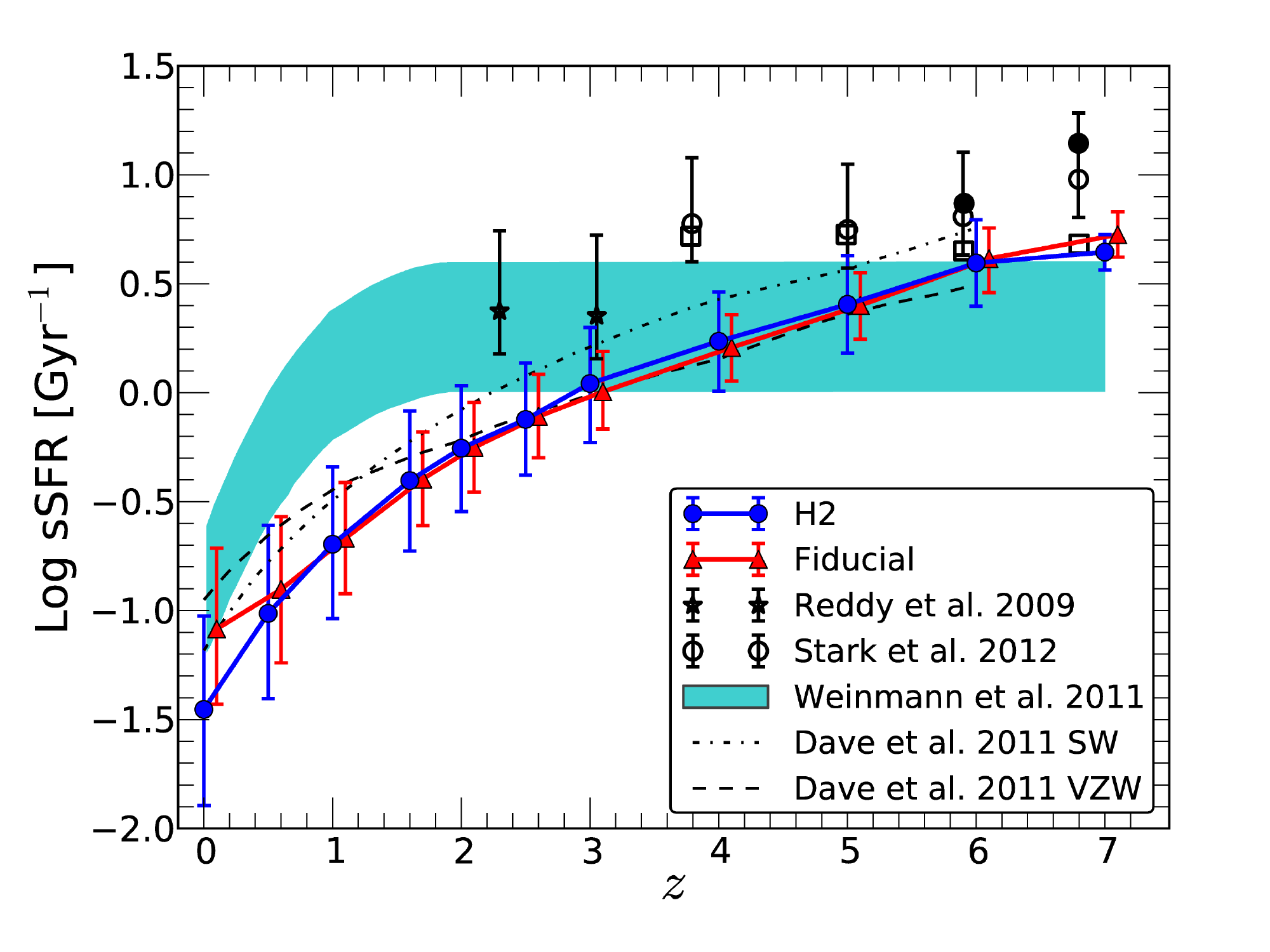}
\caption{
Redshift evolution of the sSFR of simulated galaxies.  Data points for the Fiducial and $\htwo$ runs are the median sSFR at $\mstar = 10^{10} \msun$ (Figure~\ref{fig:mstarsSFR}), while the error bars represent a $1\sigma$ spread in the data.
Observations are taken from \citet[][stars]{Reddy09}, \citet[][cyan shade]{Weinmann11}, \citet[][squares]{Bouwens12}, and \citet[][circles]{Stark12}.
Simulation data from \citet{Dave11} is shown as the black dashed (VZW model) and dot-dashed (SW model) lines for comparison.
Fiducial points are offset by 0.1 dex for clarity.
}
\label{fig:sSFRz}
\end{figure}

\subsubsection{Specific star formation rates of galaxies}
\label{sec:ssfr}

Figure~\ref{fig:mstarsSFR} shows the redshift evolution of the specific star formation rates (i.e., SFR per unit stellar mass, ${\rm sSFR \equiv SFR / \mstar}$) in our simulations.  
This plot shows the instantaneous efficiency of SF, whereas the SHMR reflects all past history of SF and feedback. 
Panel~(a) shows that the low mass galaxies in the $\htwo$ run at $z=6$ have higher sSFRs than those in the Fiducial runs, mirroring the steeper slope of the SFRD (Figure~\ref{fig:multimadau}) for the $\htwo$ run.
Our simulation data is higher than the observational data of Lyman-$\alpha$ emitters at $z=5.7$ \& 6.6 \citep{Ono10} and $z$-dropout galaxies at $z\sim7$ \citep{Labbe10}, but within their error bars. 

The $\htwo$ run in Panel~(b) ($z=3$) again show a slightly higher sSFR than the Fiducial run for 
lower mass galaxies with $\mstar \lesssim 10^{9.6}\msun$. At higher masses, the two runs agree very well,
as well as with the observed data at $z=3.7$ \& 2.0, indicated by the shaded region \citep{Daddi07,Lee11}.  
Panel~(c) ($z=1$) also shows similarly good agreement between the two runs and the observational data range  \citep{Elbaz07}.

Panel~(d) ($z=0$) shows that the sSFR of both runs continue to decrease with time, but the $\htwo$ run decreases at a faster rate. Therefore the Fiducial run has a higher sSFR than the $\htwo$ model at $z=0$. 
Both models agree with the observational data \citep{Brinchmann04} 
with a slightly decreasing sSFRD as a function of $\mstar$.

\subsubsection{Redshift evolution of the sSFR}	
\label{sec:sSFRz}

Observations indicate that galaxies of similar mass ($\sim10^{10}\msun$) have relatively constant sSFRs on the order of $1-2\,\rm{Gyr}^{-1}$ in the redshift range of $z=2-7$ \citep[e.g.,][]{Stark09,Gonzalez10,Labbe10,Labbe10b}.
This sSFR `plateau' is difficult to produce with current models of galaxy formation \citep[e.g.,][]{Bouche10, Weinmann11}. 
Figure~\ref{fig:sSFRz} shows the sSFR as a function of redshift for simulated galaxies with  $\mstar = 10^{10}\msun$.  We find that the sSFR for these galaxies decline gradually rather than a steep drop off around $z\sim1$ as observations suggest.
The $\htwo$ run however, produces a slightly steeper drop off at $z<1$ than the Fiducial run, but the normalization is still lower than the compilation of observed data points by \citet{Weinmann11}.  Neither model produces the observed plateau in the redshift range of $z=2-7$.

For comparison, simulation data from two different wind models of \citet[][dashed \& dot dashed lines]{Dave11} are included, and they show similar trends to our simulation data.
Our results are also very similar to the results of the semi-analytic model of \citet{Neistein10, Weinmann11} without the plateau at $z>2$. 
The general agreement between multiple different simulations and semi-analytic models of galaxy formation suggest that the $\Lambda$CDM model predicts a general decline in the sSFR of galaxies of a given mass, contrary to observations.  However, we note that none of these simulations included the effect of AGN feedback. \citet{Krumholz11} 
argued that taking the metallicity-dependence of $\htwo$ formation would help to reconcile the discrepancy, however, even with our new $\htwo$-based SF model, our simulations do not produce the plateau of sSFR at $z>2$. 

\begin{figure*}
\includegraphics*[width=1\textwidth,clip]{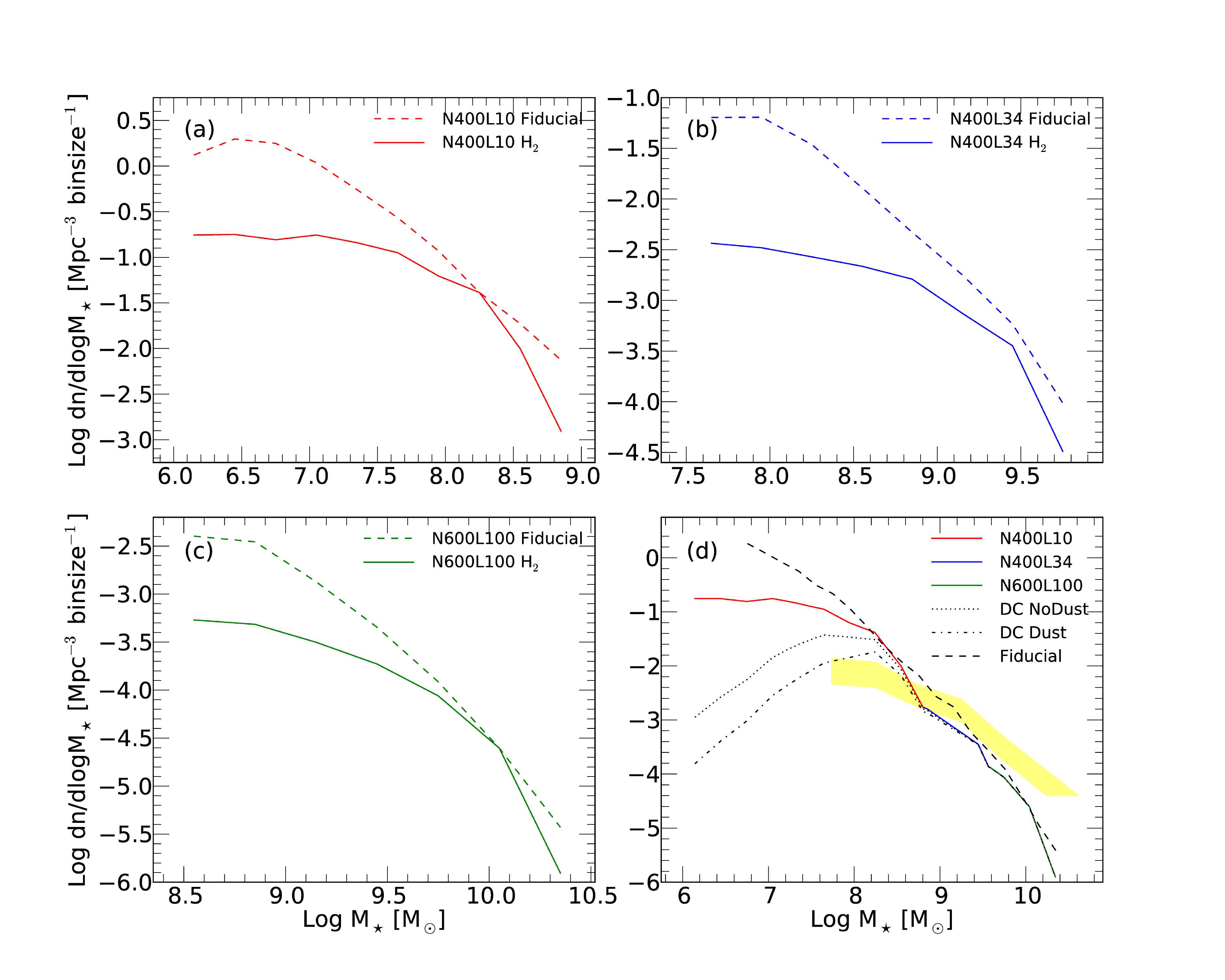}
\caption{
{\it Panels~(a-c)}: Galaxy stellar mass function for three different box sizes at $z=6$, plotted against their respective Fiducial run.  
{\it Panel~(d)}: Composite GSMF of the $\htwo$ runs, compared with the Fiducial composite GSMF (black dashed line).  Additionally we show DC corrections \citep{Jaacks12DC} with (dot-dashed) and without (dotted) dust correction.
The yellow shaded region is the observational estimate from \citet{Gonzalez10}.  
}
\label{fig:galmassfunctions}
\end{figure*}

\subsection{Galaxy stellar mass function (GSMF)}
\label{sec:GSMF}

In the previous sections, we have seen that SF is less efficient in the $\htwo$ run, which should also be reflected in the GSMF. 
Recall that for a given $\mstar$ at high-$z$, the galaxies reside in more massive halos in the $\htwo$ runs (Figure~\ref{fig:mstarVSmhalo}).  
Since the higher mass halos are less abundant in a CDM universe, this will reduce the number of low-mass galaxies and shifts the galaxy population to higher mass DM halos.

Figure~\ref{fig:galmassfunctions} shows the GSMF for our three primary runs (N400L10, N400L34, N600L100) at $z=6$.
In Panels~(a-c) we directly compare the $\htwo$ run to the corresponding Fiducial run for each simulation, and find that the $\htwo$ run produces far fewer low-mass galaxies as expected.  Note the different y-axis ranges in Panels~(a-c). 
Our result is in general agreement with the findings of \citet{Kuhlen12}; they also found a decrease in their GSMF at $
\mstar <10^9 \msun$ at $z=4$. 

Figure~\ref{fig:galmassfunctions}d shows the comparison of the composite GSMF from the two runs, 
following the method of \citet{Jaacks12}; we connect the GSMF from runs with different box sizes at the resolution limit of each run.  This method allows us to cover a wider range of $\mstar$ utilizing many simulations, and present the results collectively. 
The observational estimate from \citet[][yellow shade]{Gonzalez10} at $z=6$ is also shown. 
At the high-mass end of $\mstar > 10^9\msun$, the two composite GSMFs from $\htwo$ and Fiducial runs agree well. 
The slight kink in the composite GSMF at $\mstar \sim 10^{8.8}\msun$ for the $\htwo$ run is due to the resolution gap between the simulations; we have verified that an intermediate resolution run (N500L34, $\epsilon=2.72\,h^{-1}\rm{kpc}$) 
fills in this gap.  Due to the heavy computational load, we did not complete the corresponding Fiducial run for N500L34, therefore this run is not used for other comparisons in this paper. 
At the low-mass end of $\mstar < 10^8\msun$, the $\htwo$ run has a significantly lower number density of galaxies than the Fiducial run. 
This illustrates that the $\htwo$ model has a greater impact on the number density of low-mass galaxies.

\subsubsection{On the over prediction of GSMF at high-$z$}
\label{sec:overpredict}

One of the primary motivations for implementing the $\htwo$-based SF model was to see if it can remedy the overprediction of GSMF at low-mass end due to its natural dependence on metallicity as we described in Section~\ref{sec:intro}.
In the earlier sections, we saw that indeed the $\htwo$-based SF model reduces the number of low-mass galaxies.  However, even with the new $\htwo$ model, 
we are still over-predicting the number of low-mass objects at $\mstar = 10^{7.8}-10^{8.6}\msun$ compared to the observational estimate of \citet{Gonzalez10} at $z=6$ (Panel [d]). 
Therefore the $\htwo$ model alone does not seem to be able to solve this generic problem of CDM model. 
Our simulations also seem to under-predict the number of massive galaxies with $\mstar >10^{9.5}\msun$ when compared to the \citet{Gonzalez10} observational data  at $z=6$. 
\citet{Jaacks12} argued that this difference likely originates from the different slope in the $\mstar - $SFR relation, where the observational estimate was derived by using a crude relation from $z\sim 4$ and applied to $z=6$ assuming that it is unchanged.  In our simulations, the $\mstar - $SFR relation has a different slope, and this results in a different slope in the GSMF.

Figure~\ref{fig:galmassfunctions}d also contains the results of applying the duty cycle (DC) corrections  \citep{Jaacks12DC} to our composite GSMF both with (dot-dashed line) and without (dotted line) accounting for dust extinction. \citet{Jaacks12DC} defined the DC as the fraction of time that a galaxy exceeds the current HST magnitude limit within a certain $\Delta z$, and characterized it with a sigmoid function as a function of $\mstar$.  
According to their result, DC for $z=6$ makes a relatively sharp transition from nearly zero at $\mstar < 10^7 \msun$, crosses 0.5 at $\mstar \sim 10^8 \msun$, and to almost unity at $\mstar > 10^9\msun$. 
Using this relation, we can apply a correction for the observability of low-mass galaxies, and see the impact of SF duty cycle on the observed GSMF. 
Similarly to the results of \citet{Jaacks12DC}, our GSMF becomes closer to the observational estimate after the DC correction. 
This comparison demonstrates that the $\htwo$-SF model alleviates some of the tension between observation and simulation without having to invoke stronger stellar feedback prescriptions.

\subsubsection{GSMF at $z=3$ and 0}
\label{sec:GSMF-lowz}

\begin{figure}
\includegraphics*[scale=0.43]{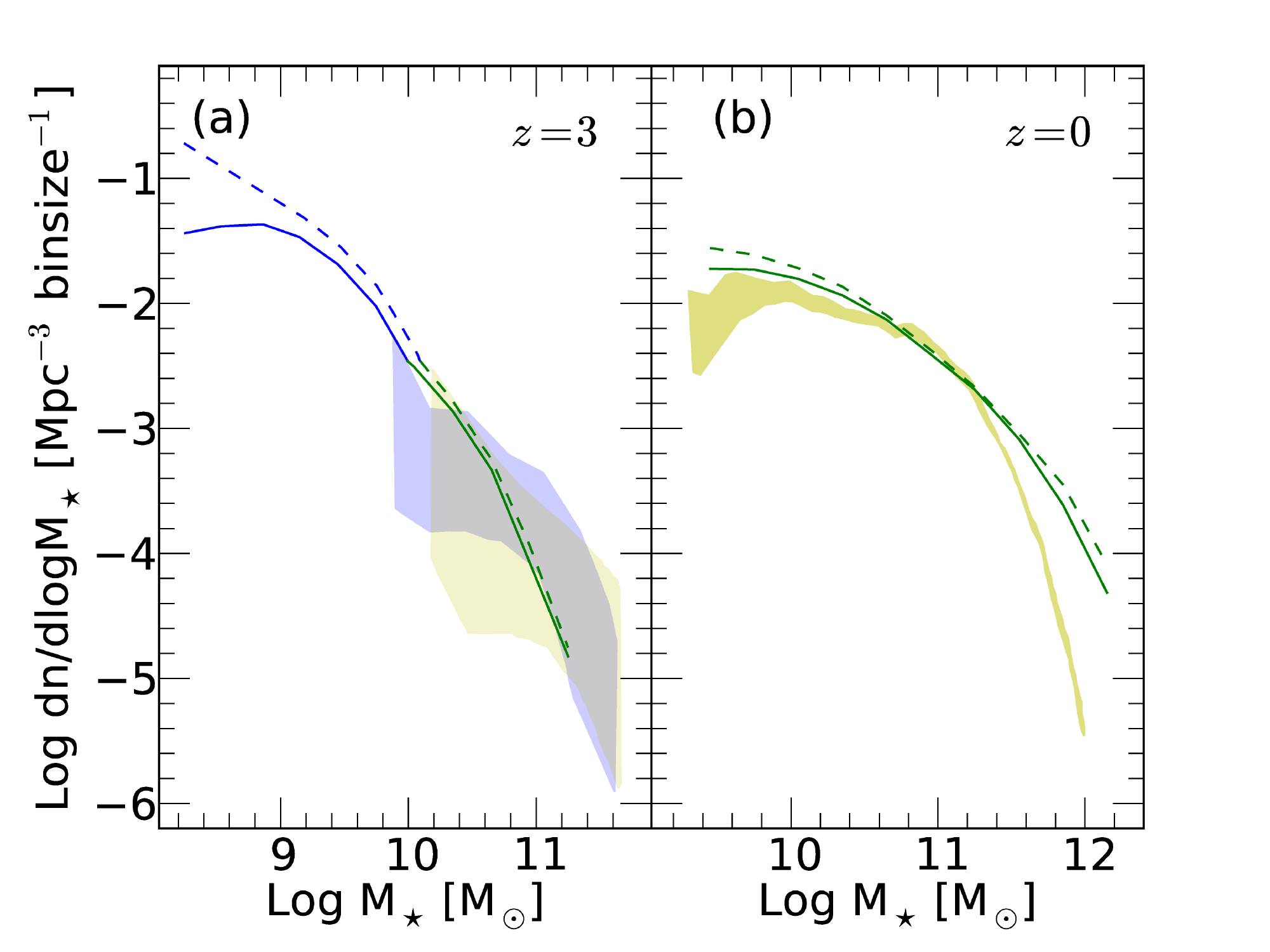}
\caption{
{\it Panel~(a)}: composite GSMF for $z=3$ for both the $\htwo$ (solid lines) and Fiducial (dashed lines) runs.  
The data is from the N400L34 \& N600L100 runs. 
The shaded regions represent observational estimates at $3<z<4$ (yellow) and $2<z<3$ (cyan) from \citet{Marchesini09}.
{\it Panel~(b)}: GSMF at $z=0$ from the N600L100 run.  The yellow shaded region is the observation from \citet{Cole01}.
}
\label{fig:galmassfunctionsz3z0}
\end{figure}

Figure~\ref{fig:galmassfunctionsz3z0} shows the GSMF at $z=3$ (Panel a) \& $z=0$ (Panel b).  
Panel~(a) is composed of data from the N400L34 \& N600L100 runs, and Panel~(b) of N600L100 data.  
Dashed lines represent the Fiducial run, while solid lines represent the $\htwo$ run.  
The shaded regions at $z=3$ represent observational estimates of the GSMF at $3<z<4$ (yellow) and $2<z<3$ (cyan) from \citet{Marchesini09}, following \citet{Choi10}.
Both sets of simulations are in agreement with each other and with observations at $\mstar \gtrsim 10^{10}\msun$, which corresponds to $\mdm \gtrsim 10^{11.5}\msun$ (Figure~\ref{fig:mstarVSmhalo}b).  
A substantial difference between the two SF models is again seen in galaxies with 
$\mstar \lesssim 10^{10}\msun$, but this is below the current observable flux limit. 

We may try to understand the discrepancies in the GSMF in relation to the SHMR.   
The difficulty is that the SHMR is not per unit volume, hence there is no obvious direct link between SHMR and GSMF.  
Suppose $\mstar$ in low-mass halos is increased uniformly, then the normalization of SHMR shifts upwards. 
At the same time, those galaxies would move from the low-mass bin to higher mass bins, and the GSMF would be weighted more towards higher mass side. 
For example, Figure~\ref{fig:SMHM}b suggests that we are producing roughly correct amount of stars in halos with $\mdm \lesssim 10^{12}\msun$ at $z=3$, and the agreement in the GSMF is not so bad either as shown in Figure~\ref{fig:galmassfunctionsz3z0}a.   Such a comparison provides a consistency check between SHMR and GSMF.

The shaded region at $z=0$ (Panel [b]) is the observational estimate from \citet{Cole01}. 
Our simulations agree well with the observation near the knee of GSMF ($\mstar \sim 10^{10.8} - 10^{11.3}\msun$), but over-predicts at both low and high mass end.  
This over-estimation at $\mstar > 10^{11.3}\msun$ is reflected in the overestimation of the SHMR at $\mtot \sim 10^{13}\msun$ (Figure~\ref{fig:SMHM}d), which could be due to a lack of AGN feedback in our current simulations. 
At the low-mass end ($\mstar \lesssim 10^{10.5}\msun$), both models over-predict the GSMF, but the $\htwo$ run to a lesser degree.  

It is clear that simultaneously matching the SHMR and GSMF is not an easy task.  We expect the inclusion of AGN feedback will assist in improving the high-mass end of our simulations at low redshift.  The new $\htwo$-based SF model seems to have improved the relations in regards to the low-mass end, but does not fully reconcile the differences.  Further improvements to our SN feedback prescriptions (e.g., momentum feedback by winds) may be required to achieve better agreement with observations. 


\subsection{Kennicutt-Schmidt (KS) relationship}
\label{sec:KSlaw}

\begin{center}
\begin{figure*}
\includegraphics*[width=1.0\textwidth,clip,scale=0.5]{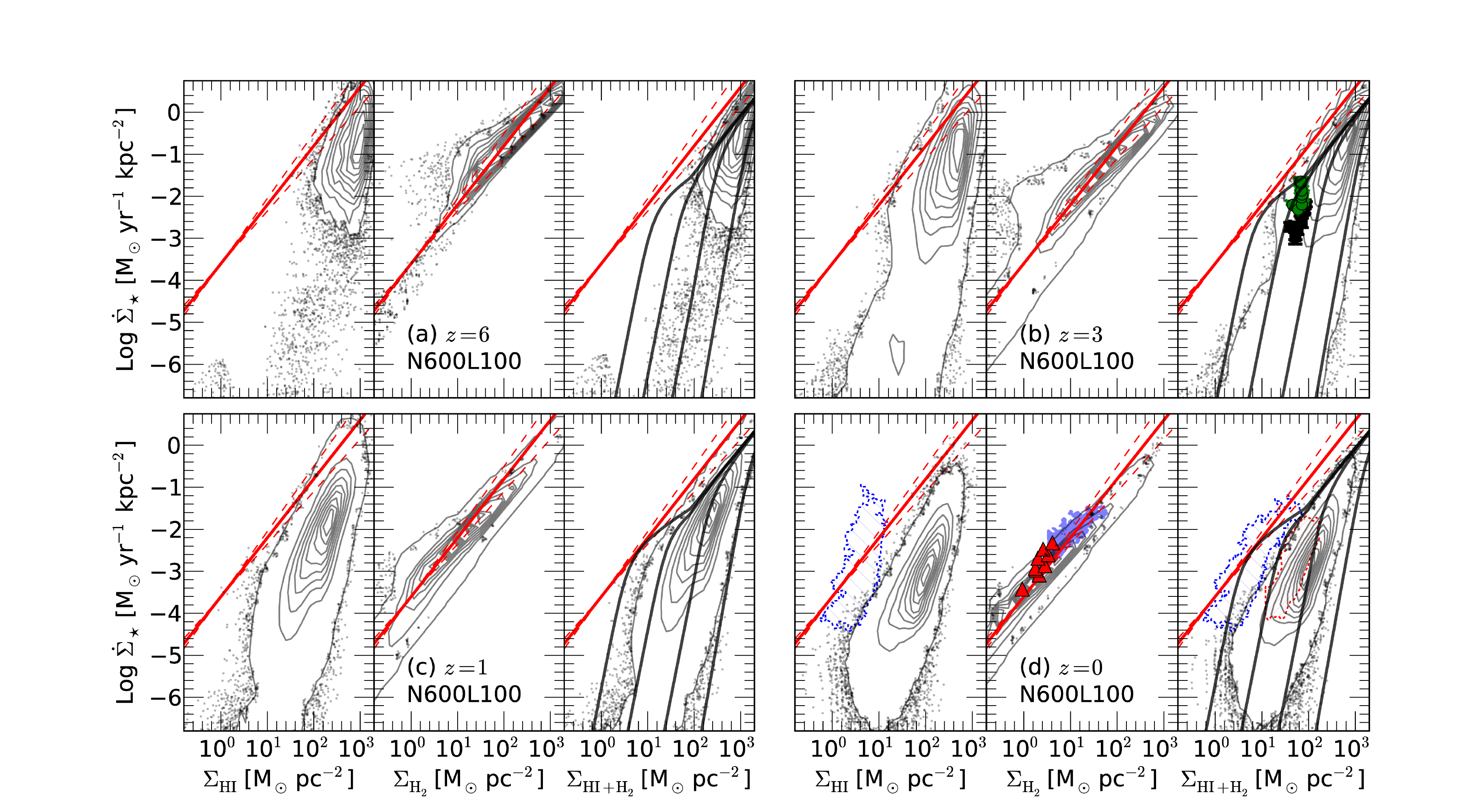}
\caption{
The Kennicutt-Schmidt relation for the N600L100 simulation at $z=6,3,1$, \& 0.  Each redshift is broken into three panels: SFR surface density as a function of \HI (left panel, hereafter KS-HI), $\htwo$ (middle panel, KS-H2), and \HI\ +$\htwo$ surface density (right panel, KS-HIH2).
In each panel, the solid red line represents the empirical KS relation given by Equation~(\ref{eq:kslaw}), and the dashed red lines represent the range of slope $\Delta n=\pm0.15$.
Blue hatched regions in the KS-HI(d) and KS-HIH2(d) panels, and the blue solid contour in KS-H2(d) panel are the observations from \citet{Bigiel08}.
The red triangles in the KS-H2(d) panel along with the red hatched region in the KS-HIH2(d) panel are low-metallicity SMC observations from \citet{Bolatto11}. 
At $z=3$ we have observational data from \citet{Rafelski11} plotted as green circles and black squares.  The green circles represent upper limits derived for DLAs, while the black squares represent outskirts of LBGs.
Lastly in KS-HIH2 panels, black lines represent theoretical results from the KMT model \citep{Krumholz09}.
In the $z=6$ KS-HIH2 panel, the four theory lines correspond to the metallicities $\log (cZ/\Zsun)=0.11, -0.69, -1.49, -2.29$ from left to right, respectively.  For $z=3$, we have $\log (cZ/\Zsun)=0.47, -0.45, -1.37, -2.29$. For $z=1$, $\log (cZ/\Zsun)=0.77, -0.25, -1.27, -2.29$, and for $z=0$ $\log (cZ/\Zsun)=0.80, -0.23, -1.26, -2.29$.
Any discreteness of the dotted points at the contour edge is an artifact from our plotting procedure.
}
\label{fig:kennicuttbreakdown}
\end{figure*}
\end{center}

Ideally we would like to reproduce the empirical Kennicutt-Schmidt relationship naturally in simulations;   previously the KS relation was imposed in our SF prescription \citep{Choi10} and in many others', therefore the results matched the observation well by construction.  The new $\htwo$-based SF model provides two main benefits:  it is not `tweaked' to match the KS relation, and it is more physically realized in that stars are formed out of cold molecular gas on a depletion time-scale which is equal to about 1\% of the free-fall time (i.e., with 1\% efficiency per free-fall time). 

To examine the KS relation, we calculate the column density of \HI, $\htwo$, and SFRs along the z-axis of each halo in our simulation on a uniform grid with a cell size of $\epsilon^{2}$.
A detailed description of this process can be found in \citet{Nagamine04column}.
In Section~\ref{sec:H2regsf}, we stated that the $\htwo$ model was accurate for $Z \geq10^{-2} \Zsun$, yet we set our metal floor below that at $Z =10^{-3} \Zsun$.  
The model fails at low metallicities by over-predicting the amount of $\htwo$ mass.  This is due to time-dependent effects being neglected within the analytical KMT model \citep{KrumholzGnedin11}.
However, the over-predicted value may be an accurate estimate of how much cold material is present to form stars \citep{Krumholz12}.
Therefore we simply assume that the $\fhtwo$ value calculated by Equation~(\ref{eq:fh2calc}) for
any gas particle with $Z <10^{-2} \Zsun$ is actually representative of the amount of cold \HI\ gas, which is available for star formation. 

Figure~\ref{fig:kennicuttbreakdown}(a-d) shows the KS relation for the N600L100 simulation at $z=6,3,1$, \& 0.
Each point in this figure represents one pixel on the projected x-y plane, and the contour is used to represent all the columns from all halos in the simulation box. 
For each redshift, the panel is broken down into three sub-panels: 
the first being the KS relation for \HI\ gas only, second is for $\htwo$ gas only, and lastly for \HI\,+$\htwo$.
We will refer to these panels as KS-HI, KS-H2, and KS-HIH2, respectively.
Each panel includes the KS relation given by Equation~(\ref{eq:kslaw}) as a solid red line, with the dashed lines representing the range of slope $\Delta n=\pm0.15$.

In KS-HI panel at $z=0$, we also overplot the observational data from seven nearby spiral galaxies as a blue hatched region \citep[][hereafter B08]{Bigiel08}.
In KS-H2 panel ($z=0$), we overplot the low surface density observations from the Small Magellanic Cloud (SMC) as red triangles \citep[][hereafter known as B11]{Bolatto11}.
Lastly in KS-HIH2 panel ($z=0$), we again plot B08 data as a blue hatched region, and B11 data from the SMC as a red hatched region.   

There are two major processes driving the evolution of these plots.
The first is gas depletion: as time passes the cold molecular gas used to form stars is depleted, and become less available at late times.  This is most obvious in the decrease of $\sigmasfr$ between $z=3, 1, \&\,0$, corresponding to the downturn of the SFRD at $z\lesssim 2$ in Figure~\ref{fig:multimadau}. 
The second is metal enrichment: the longer a simulation runs, the more enriched the gas becomes via SF. 
This process expands the distribution of points to the left-hand-side of the plot, because higher metal content allows stars to form at lower surface densities, as shown in Figure~\ref{fig:svalues}.
The distribution of points broadens from $z=6$ to $z=0$, indicating greater range of metallicities present in the simulations.

The KS-HIH2 panels include theoretical results from the KMT model \citep{Krumholz09} to show the same effect.  
The column densities calculated for each pixel represent the smoothed value on a relatively large projected scale of $\epsilon^2$; if we use this value, the model will underpredict $\fhtwo$, since it does not account for clumping of the gas on scales below our simulation's spatial resolution limit of $\epsilon=4.30\,h^{-1}\rm{kpc}$, as well as the path-length along the column. 
To account for this effect, the KMT model multiplies the calculated gas column density by a clumping factor ``$c$'' ($\Sigma_{\rm{HI}+\htwo}=c\times\Sigma_{\rm{calc}}$), which increases the surface densities to be compared with observations.  In order to compare the KMT model result with our simulation, the theory lines are shifted to lower `computed' surface densities (i.e., $\Sigma_{\rm calc} = \Sigma_{\rm{HI}+\htwo} / c$), which brings a good agreement between the KMT model results and our simulations.   In Figure~\ref{fig:kennicuttbreakdown}, we adopted $c=5$.

\begin{figure}
\includegraphics*[width=0.50\textwidth,clip]{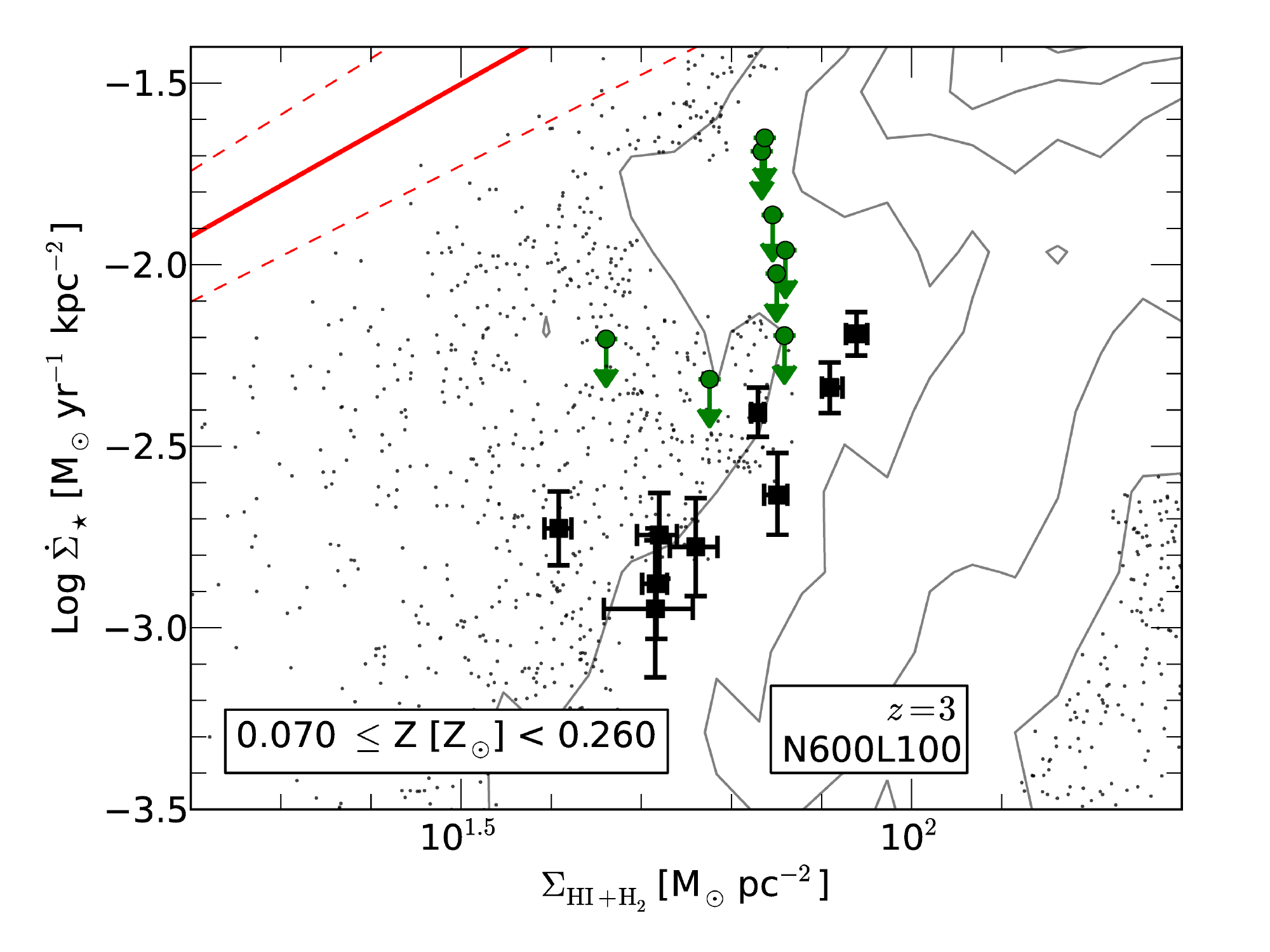}
\caption{
Enlarged region of Figure~\ref{fig:kennicuttbreakdown}b KS-HIH2 plot.  Here we only plot star-forming columns with metallicities consistent with observations from \citet{Rafelski11} ($Z=0.07-0.26\,\Zsun$).  Green circles represent upper limits derived for DLAs \citep{Wolfe06}, and black squares represent outskirts of LBGs \citep{Rafelski11}.  The observed data points are at the upper edge of the simulation data contour, but there are many simulated columns that overlap with the observed data. 
}
\label{fig:DLALBGz3}
\end{figure}

For the KS-HI panel at $z=0$, we find disagreement between simulation and the B08 data (blue hatched region).
This is a metallicity effect; our simulations do not contain enough high-metallicity columns, and the low metallicity columns will form stars at higher surface densities in the KMT model.

In the KS-H2 panel at $z=0$, our simulation is in good agreement with the observations.
The $\sigmasfr$ starts high at $z=6$, and eases its way to the lower left due to the two processes described above.  
By $z=0$, the observations lie in the center of our simulation data showing a very good agreement even for low surface densities.
It should be noted that we are directly measuring the amount of $\htwo$ in our simulation, whereas the observers infer this value from the CO luminosity. 

In the KS-HIH2 panel at $z=0$, we again find a disagreement between simulation and the B08 data (blue hatched region); the bulk of our data is found at slightly higher surface densities compared to these observations.  The data in these panels is dominated by \HI, resulting in similar trends to the KS-HI panel.

In the KS-HIH2 panel at $z=3$ (Panel [b]), we also overlay the upper limits from damped Lyman-alpha absorbers (DLAs) as green circles and outskirts of Lyman-break galaxies (LBGs) as black squares \citep{Wolfe06,Rafelski11}.
LBGs are considered to be star-forming galaxies with moderate median mass of $\mstar \sim 10^{10}\msun$, 
therefore are expected to have been enriched to some level.
\citet{Rafelski11} find the LBGs in their sample have  $Z\approx 0.07-0.26\,\Zsun$.  
Figure~\ref{fig:DLALBGz3} shows the KS plot for only star-forming columns in our N600L100 simulation with 
$Z=0.07-0.26\,\Zsun$ at $z=3$.
The observed data points are close to the edge of the simulation contour, but there are many columns that agree with the observational data.  Note that it is certainly easier to observe the SFR closer to the upper edge of the contour rather than the bottom side of it due to the surface brightness dimming.

\begin{figure*}
\includegraphics*[width=1\textwidth,clip,scale=0.5]{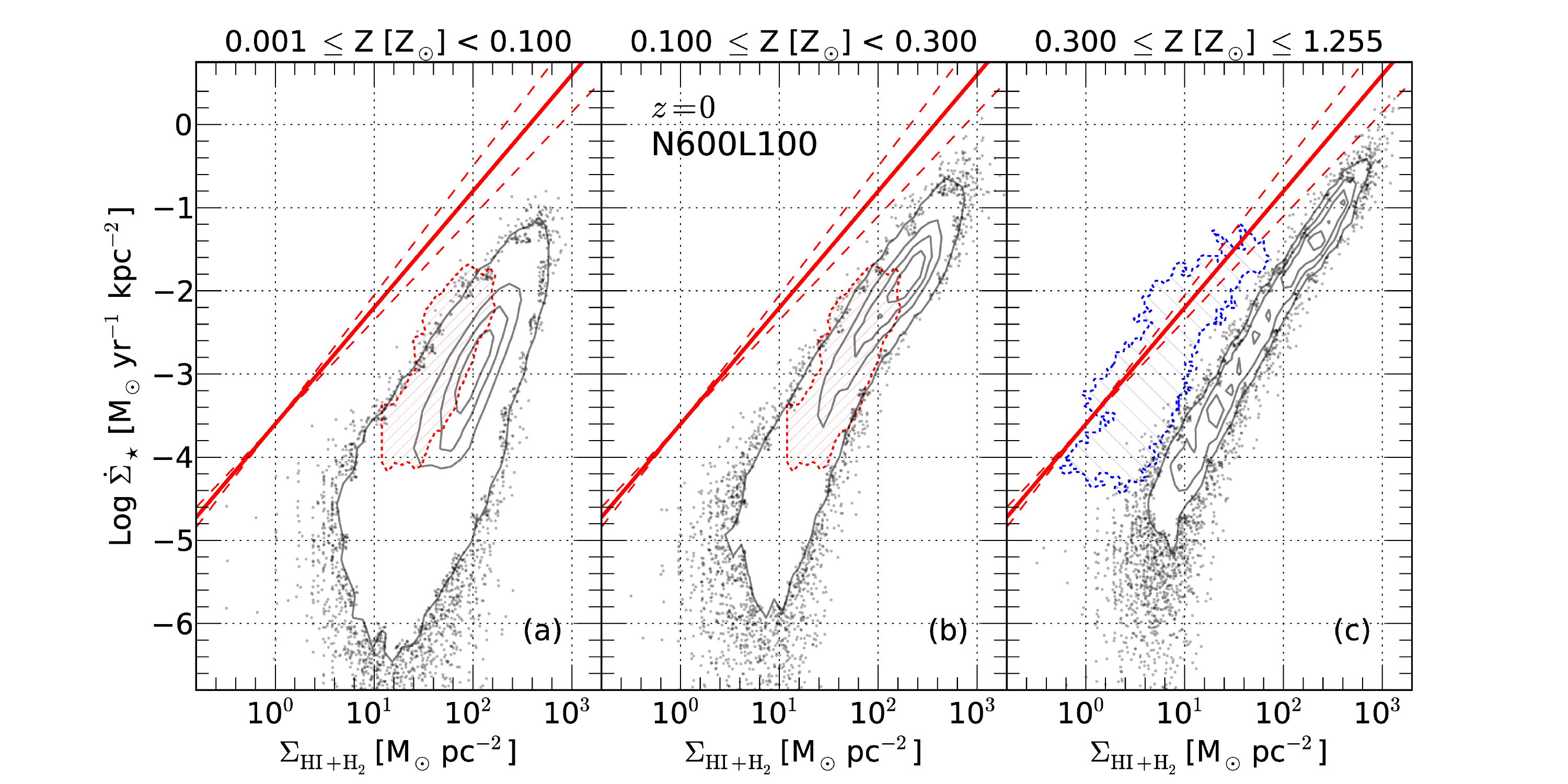}
\caption{
The KS-HIH2 panel from Figure~\ref{fig:kennicuttbreakdown}d is separated into three different metallicity ranges at $z=0$ using the N600L100 run, in order to show the metallicity effect on the KS plot.
Panel~(a) only shows columns with the lowest metallicities.  The metallicity range in panel~(b) brackets the \citet{Bolatto11} SMC data.  Panel~(c) shows the highest metal columns, however, the median metallicity of the simulated columns are biased towards the lower end of the bracket $Z=0.3\,\Zsun$, which is presumably causing the offset between the simulation result and the blue hatched observed data. 
}
\label{fig:kennicuttmetal}
\end{figure*}

Figure~\ref{fig:kennicuttmetal} further illustrates the metallicity effect by separating the KS-HIH2 panel from Figure~\ref{fig:kennicuttbreakdown}d into three different metallicity ranges for the N600L100 run at $z=0$. 
We find that the columns with the lowest metallicity (Panel [a]) are forming stars at the highest gas surface densities for a given $\sigmasfr$ as expected.
Panel~(b) brackets $Z = 0.2\,\Zsun$, which is roughly equal to the metallicity of the SMC \citep{Bolatto11}.  Columns in our simulation in this metallicity range agree very well with the observed data (red contour).
In Panel~(c) we show columns of higher metallicity $Z>0.3\,\Zsun$, which is similar to the range of B08 sample ($0.41 - 0.69\,\Zsun$) \citep{Walter08}.  
As discussed previously, our data does not agree very well with observations in this range.  There are some points overlapping with the observed data, but the majority lie at higher column densities than the observed range. 
This discrepancy is presumably caused by different metallicities:
the highest metal column at $z=0$ in the N600L100 run is $Z = 1.26\,\Zsun$, yet the median 
column metallicity at $Z >0.3\,\Zsun$ is $Z = 0.41\,\Zsun$.
This suggests that our N600L100 run does not contain enough high metallicity columns to match these observations. If the simulation had more high metallicity columns, then the SF would occur more at lower gas surface density, and there would be more points overlapping with the B08 data. 
This discrepancy could be related to our SN feedback and galactic wind models since the metal recycling process is tightly connected with them. 

\subsection{\HI\ \& H$_2$ column density distribution functions}
\label{sec:neutralhydrogen}

One of the best ways to investigate the distribution of \HI\ gas in the Universe statistically is to examine the \HI\ column density distribution function $\fHI$ \citep[e.g.,][]{Nagamine04sf, Nagamine04column,Wolfe05, Zwaan06, Prochaska09,Noterdaeme09,Pontzen10, Altay11, Yajima11,  Rahmati12,Erkal12,Noterdaeme12}.
Using the Fiducial runs, \citet{Nagamine10} found that a simple self-shielding model with a threshold density ($n_{\rm th}^{\rm UV}=6\times10^{-3}$\,cm$^{-3}$) for UVB penetration can reproduce the observed $\fHI$ quite well at $\log \NHI < 21.5$ for $z=3$. 
\citet{Yajima11} later showed the validity of $n_{\rm th}^{\rm UV}$ value using full radiative transfer calculations. 
However, \citet{Nagamine10} also found that the Fiducial run over-predicts  $\fHI$ at $\log \NHI>22$, and argued that this might be due to the neglect of $\htwo$ within the Pressure SF model (Section~\ref{sec:pressuremodel}), because then part of \HI\ would be converted into $\htwo$ and $\fHI$ would decrease at high $\NHI$ values. 

\begin{figure}
\includegraphics*[scale=0.36]{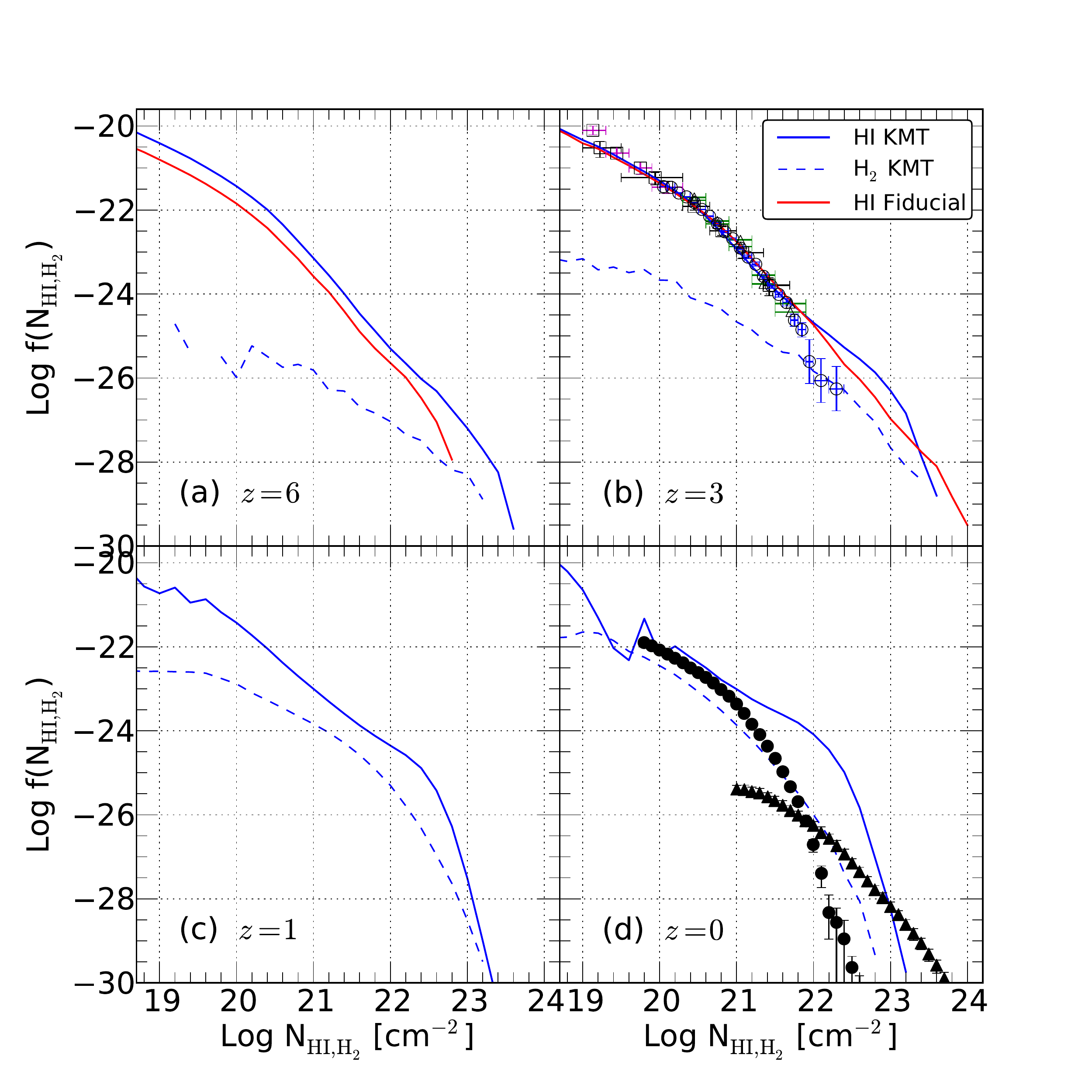}
\caption{
Column density distribution functions of \HI\ and $\htwo$ at $z=6, 3, 1,$ \& 0 for the $\htwo$ and Fiducial runs.  Redshifts $z=6$ \& 3 are from the N144L10 runs, while $z=1$ \& 0 are from the N600L100 $\htwo$ runs (N600L100 Fiducial was omitted because it lacks the OTUV threshold).
The observational data points at $z=3$ are from \citet[][black squares]{Peroux05}, \citet[][magenta squares]{Omeara07}, \citet[][green triangles]{Prochaska09}, and \citet[][blue circles]{Noterdaeme12}.
Panel~(d) shows observations from \citet{Zwaan06}, where black circles represent the \HI\, and black triangles represent the $\htwo$ column density distribution functions.
}
\label{fig:neutralcolumn}
\end{figure}

Figure~\ref{fig:neutralcolumn} compares the column density distributions of both \HI\ and $\htwo$ in the $\htwo$ and Fiducial runs at $z=6, 3, 1, \&\, 0$.
Panels~(a) \& (b) are composed of N144L10 data, while Panels~(c) \& (d) are composed of N600L100 data. 
The Fiducial run is omitted from panels (c) \& (d), because the N600L100 Fiducial run did not use the OTUV threshold which is necessary to bring the column density distribution into agreement with observations at $z=3$ \citep{Nagamine10}.  

In Panel~(a) we see that the $\htwo$ run consistently has higher amplitude of $\fHI$ than the Fiducial run due to less efficient star formation. 
At $z=3$ (b) however, we find that the $\htwo$ run has a higher $\fHI$ than that of Fiducial run at $\log \NHI > 22$. 
This is because the star formation is less efficient in the new $\htwo$ run, therefore more \HI\ gas is left over in high density regions.  In the $\htwo$ run the varying SF threshold density was higher than the constant $n_{\rm th}^{\rm SF}$ adopted in the Fiducial run (Figure~\ref{fig:rhothresh}), and it was also clear from Figure~\ref{fig:phasecompare} that the gas particles are reaching higher densities in the $\htwo$ run before being heated by SN feedback than in the Fiducial run. 
The $\fHI$ results at the lower $\NHI$ do not change between the two runs at this redshift.
Panels~(c) \& (d) continue to show the redshift evolution of this relationship in our simulations.  
At $z=0$, we find that our simulations over-predict $\fHI$ at $\log \NHI>21$, over-predict the $f(H_2)$ at $\log N_{\rm H_2}<21$, and under-predict at $\log N_{\rm{H}_2}>22$.

Therefore the current simulations suggest that it is difficult to explain the sharp turn-down of observed $f(\NHI)$ at $\log \NHI \sim 22$ by the atomic-to-molecular transition, in agreement with the conclusions of \citet{Erkal12}.
Additionally, \citet{Erkal12} showed that their simulations could be brought into agreement with observations if a region of 3 kpc radius around the center of all galaxies was removed.  This could be another opportunity for AGN feedback to play an important role: if feedback from super massive black holes can prevent the formation of high columns, then our simulations may come into better agreement with observations at $N>10^{22}\rm{cm}^{-2}$.  Obviously more refinement of feedback models are needed to bring the simulations into agreement with the observations of $\fHI$ and $f(N_{\rm H_2})$.

\subsection{Resolution studies}
\label{sec:resdepend}

The new $\htwo$-based SF model has an implicit resolution dependence.
With higher resolution, the simulation resolves higher (column) densities (Eq.~\ref{eq:sobolev}), which yield lower $s$ values (Eq.~\ref{eq:sval}) for a given metallicity.
Figure~\ref{fig:svalues} illustrates that lower $s$ values lead to higher $\fhtwo$ (Eq.~\ref{eq:fh2calc}), which increases the SFR (Eq.~\ref{eq:h2sfr}).

To examine the resolution effect, Figure~\ref{fig:kennicuttres} shows the KS relation for the N600L10, N400L10, N400L34, \& N600L100 runs at $z=6$.  These panels are ordered by resolution: Panel~(a) shows the highest resolution, and Panel~(d) shows our lowest resolution simulation.
In Panels~(a) \& (b), we can examine the resolution effect on the KS plot when keeping the box size constant.  In general the gas surface densities where SF takes place do not change very much, but with higher resolution, the points cover a wider range of $\sigmasfr$. 
This is an expected result from a higher resolution simulation; the additional resolution allows the gas to collapse to higher densities, yielding additional shielding which eases the transition to $\htwo$.

While Panels~(a) \& (b) are both from simulations with a box size of comoving $10h^{-1}$\,Mpc, Panels~(c) \& (d) are from simulation boxes of $34h^{-1}\rm{Mpc}$ and $100h^{-1}\rm{Mpc}$, respectively.
Increasing the box size of a simulation usually comes with a price of decreasing the resolution, and it 
results in more higher mass halos and fewer low-mass halos \citep[e.g.,][]{Thompson12}.  Note that the simulations shown in Panels~(c) \& (d) are of lower resolution than those in Panels (a) \& (b). 
When comparing the L10 boxes with the L34 \& L100 boxes, we are actually examining the resolution and box size effects simultaneously.  
Comparing all the panels in Figure~\ref{fig:kennicuttres} suggests that our KS results are not significantly affected by these resolution effects.  The only visible effect we see in the figure is that the lower resolution results in a thinner contour distribution

\subsubsection{Probability Distribution Function (PDF) of $\htwo$ density}
 
\begin{figure}
\includegraphics*[scale=0.43]{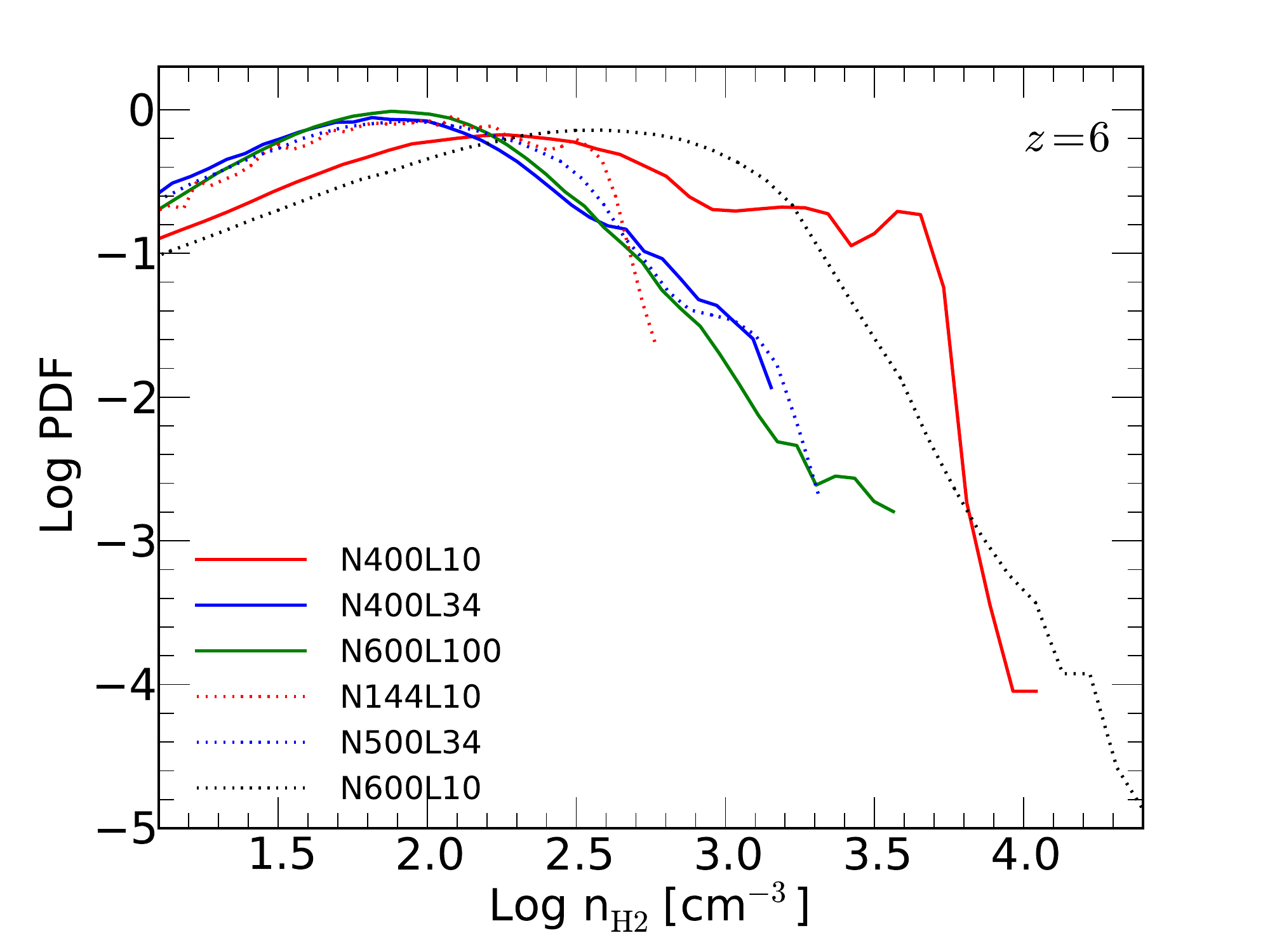}
\caption{
Probability distribution functions of physical $\htwo$ number density for all runs in Table~\ref{table:sims} at $z=6$.   
One can see that the peak of the highest density region shifts to higher densities as the resolution increases.}
\label{fig:pdfhighdensity}
\end{figure}

Physical number densities of observed molecular clouds are on the order of a few hundred cm$^{-3}$, in rough agreement with the highest densities achieved in our current cosmological simulations.
Figure~\ref{fig:pdfhighdensity} shows the mass-weighted PDF of $\htwo$ number density at the highest densities in our simulations at $z=6$. 
As expected, we can see that the peak of the highest density region shifts to higher densities as the resolution increases; 
the lowest resolution production run (N600L100) has a peak at $n_{\htwo} \sim 10^{2}$\,cm$^{-3}$, and 
our highest resolution production run (N400L10) has a peak at $n_{\htwo} \sim 10^{3.6}$\,cm$^{-3}$. 
However, the N400L10 run has a slightly different shape from the other runs, and the higher resolution N600L10 run has a peak at a slightly lower value of $n_{\htwo} \sim 10^{2.8}$\,cm$^{-3}$.
The exact reason for this different PDF shape is unclear, but presumably it was affected by some SF events. 

Earlier, \citet{Jaacks12} showed that the Fiducial runs do not satisfy the \citet{BB97} mass resolution criteria, 
even though gas particles in our N400L10 have particle masses lower than the typical Jeans mass at $z=6$ by a factor of $\approx1-100$. 
This prevents us from explicitly resolving the collapse of star forming molecular clouds directly, and it is one of the primary reasons for employing a sub-grid model for SF using the KMT model. 
Given that the highest densities reached in our simulations is approximately equivalent to those of observed giant molecular clouds, we consider that the KMT is suitable to use as a sub-grid model in our simulation to estimate the $\htwo$ mass for star formation.  In fact, the KMT model is well suited to predict the galactic-scale trends in atomic and molecular content rather than the structure of individual photo-dissociation regions \citep{Krumholz08,  Kuhlen12}.

\section{Summary}
\label{sec:summary}

We have implemented a new $\htwo$-based SF model in our cosmological SPH code \gadget.  
Previous SF models did not consider the formation of $\htwo$, and imposed the KS relation in their SF prescriptions. 
The analytic KMT model has provided a computationally inexpensive way to estimate the $\htwo$ mass fraction in cosmological hydrodynamic simulations, which allows us to modify our SF prescription to compute the SFR based on $\htwo$ density rather than total gas density. The model brings a natural dependence of star formation on metallicity (in addition to the previous dependence through metal line cooling). 

We performed a series of cosmological simulations with different box sizes and resolutions, and 
examined how this new $\htwo$-based SF model affected the results such as stellar-to-halo mass ratio, cosmic SFRD, galaxy specific star formation rates, galaxy stellar mass function, Kennicutt-Schmidt relationship, and \HI\ \& $\htwo$ column density distributions.
We find that this new $\htwo$-based SF model provides many advantages over previous models, and we summarize our primary conclusions below.

\vspace{0.2cm}
\begin{itemize}

\item In the new $\htwo$-based model, each gas particle has different SF threshold densities based on its metallicity (Figure~\ref{fig:rhothresh}).  We have shown that the new SF threshold densities (i.e., metallicity-dependent density required for $\htwo$ formation)
are higher than the constant threshold density used in the Fiducial run, which results in overall decrease of SFRD (Figure~\ref{fig:multimadau}) in the new model.  Decrease of star formation leads to weaker feedback effects subsequently. 
The need for sufficient shielding from radiation field for $\htwo$ formation results in lower SFR, causing a gas reservoir to build up.  Consequently, SF starts later than in the Fiducial run, and the peak of SFRD has slightly shifted to a lower redshift.  But both runs are still compatible with the observed range of SFRD in the Lilly-Madau diagram. 

\vspace{0.2cm}
\item The $\htwo$ run is able to successfully reproduce the SHMR at $z=3$ \& 6 for lower mass halos with $\mtot < 10^{12}\msun$ (Figure~\ref{fig:SMHM}).  
The Fiducial run with previous SF model significantly overpredicts SHMR at the same mass range, therefore the $\htwo$ run provides a significant improvement on this aspect. 
Since the SN feedback model was kept the same in the two runs, this improvement was purely driven by the change in the SF model, rather than the feedback. 

Both runs overpredict the SHMR in halos $\mtot >10^{12}\msun$ at $z<3$, which might be due to lack of AGN feedback in our current simulations.  This is connected with the over-prediction of GSMF at the high-mass end in our simulations. 

\begin{figure*}
\includegraphics*[width=1\textwidth,clip,scale=0.5]{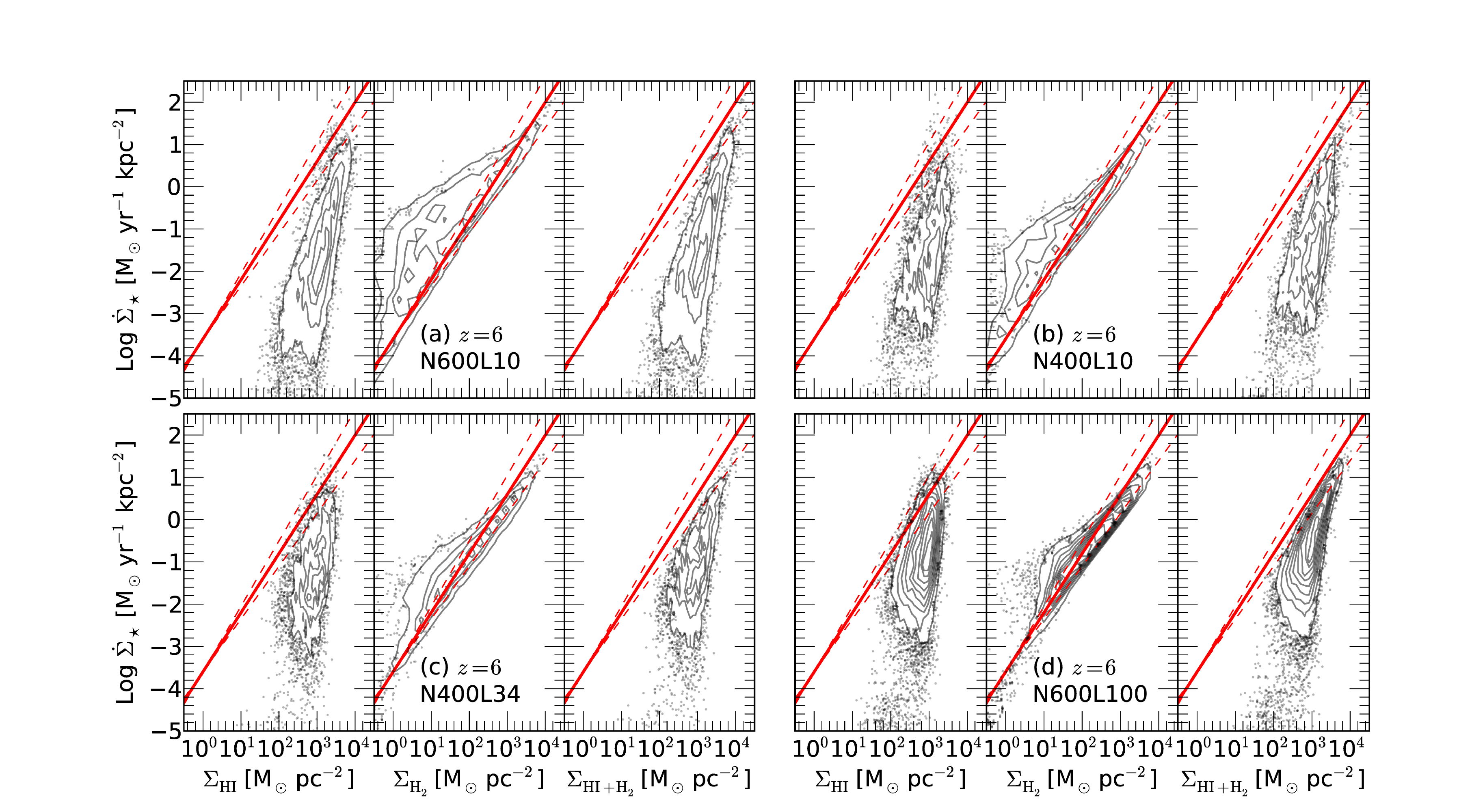}
\caption{
Similarly to Figure~\ref{fig:kennicuttbreakdown}, we plot the KS relation for the N600L10, N400L10, N400L34, \& N600L100 simulations at $z=6$ to examine the resolution and box size effects.  See text for detailed discussions. 
}
\label{fig:kennicuttres}
\end{figure*}

\vspace{0.2cm}
\item The sSFRs of galaxies in the $\htwo$ and Fiducial runs are in rough agreement with observations (Figure~\ref{fig:mstarsSFR}), and they decrease systematically with decreasing redshift.  At $z=6$, the $\htwo$ run has a higher sSFR for galaxies with $\mstar < 10^{10}\msun$, but this is due to the fact that the galaxies with same $\mstar$ reside in higher mass halos in the $\htwo$ runs than in the Fiducial run (Figure~\ref{fig:mstarVSmhalo}).   At later times, this difference becomes much smaller and the two models are in rough agreement with one another.

However, the median sSFR of simulated galaxies with $\mstar = 10^{10}\msun$ does not behave consistently with the observations as a function of redshift (Figure~\ref{fig:sSFRz}).  Both our Fiducial and $\htwo$ runs predict gradually declining sSFR with decreasing redshift, similarly to other previous models \citep{Bouche10, Dave11, Weinmann11}. 
\citet{Krumholz11} argued that taking the metallicity-dependence of $\htwo$ formation would help to reconcile the discrepancy, however, even with our new $\htwo$-based SF model, our simulations do not produce the plateau of sSFR at $z>2$. 
The general agreement between multiple different simulations and semi-analytic models of galaxy formation suggest that the $\Lambda$CDM model predicts a general decline in the sSFR of galaxies of a given mass, contrary to observations.  However, we note that none of these simulations included the effect of AGN feedback.

\vspace{0.2cm}
\item We find that the $\htwo$-based SF model produces significantly fewer galaxies at $\mstar<10^{8}\msun$ compared to the Fiducial run at $z=6$ (Figure~\ref{fig:galmassfunctions}d).  
Even after this reduction, the faint-end slope of GSMF in the $\htwo$ run is still steeper than what has been observationally estimated at $z=6$. 
Employing duty cycle corrections following \citet{Jaacks12DC} brings the GSMF closer to observations.

At $z=3$ we find that our simulations are in good agreement with observed GSMFs at $\mstar > 10^{10}\msun$, consistent with our previous finding in \citet{Choi10}. 
At the lower masses of $\mstar < 10^{10}\msun$, again the $\htwo$ model produces fewer number of low-mass galaxies relative to the Fiducial run.  At the moment, the flux limit of GSMF data is $\mstar \sim 10^{10}\msun$ even with the deepest HST imaging, and there is no good data below this limit. 
Galaxies with $\mstar < 10^{10}\msun$ correspond to halos with $\mdm < 10^{12}\msun$, and in this regime the new $\htwo$ run agrees with the observational estimate of SHMR much better than the Fiducial run. For this reason, we expect that the $\htwo$ run would match the observations of GSMF better in the future at $\mstar < 10^{10}\msun$. 

Finally at $z=0$, we find that our simulations over-predict the GSMF at both low and high-mass end. 
The deviation at the low-mass end seems smaller than at the high-mass end, however, since this is a log-log plot, the actual deviation is greater at the low-mass end. 
Further improvement in our feedback prescriptions (e.g., momentum-driven feedback by SN and AGN) may be needed to reconcile these differences.

\vspace{0.2cm}
\item We find that the new $\htwo$-based SF model naturally produces the empirical Kennicutt-Schmidt relationship without the need for `tweaking' the parameters of the SF model. 
The most significant discrepancy between the $\htwo$ run and observation can be seen against the nearby spiral galaxy data of \citet{Bigiel08}.  
It seems that the $\htwo$ run does not contain enough high-metallicity columns to match observations of nearby spiral galaxies.
These discrepancies indicate that our current simulations might have too much low-metallicity gas in massive galaxies at $z<2$, which is also indicated by Figure~\ref{fig:mstarfgas}d.
However the $\htwo$ run is able to match the observations of DLAs and LBGs at $z=3$, as well as 
the observations of the low-metallicity SMC by $z=0$.

\vspace{0.2cm}
\item As for the hydrogen column density distribution function, we find that the new $\htwo$ model did not improve the agreement with observation at $\log N_H > 21.6$ at $z=3$, and we still over-predict $\fHI$ similarly to the previous simulations. 
\citet{Erkal12} also concluded that the atomic-to-molecular transition alone could not account for the downturn in $\fHI$ at $\log N_H > 21$.
At $z=0$, our simulations do not agree with the observational data of \citet{Zwaan06}, and further refinement of star formation and feedback models are needed. 

\end{itemize}

Finally, it is important to point out that SN feedback and subsequent metal enrichment is tightly coupled with SF within our simulations.  
We must be mindful of this when comparing to observations, as changing the SF prescription could inherently change the metal enrichment process.  While the new $\htwo$-SF model may alleviate some of the discrepancies between simulation and observations, we recognize that the problem is not completely solved.  Further improvements to our feedback prescription may be required in order to reach better agreement with observations.

As for our future plan, we intend to improve our simulations on a few fronts, given the problems that we observed in this paper.  Since the $\fhtwo$ calculated in the KMT model depends on gas metallicity, we need to account for the metal diffusion in the ISM more accurately \citep[e.g.][]{Shen10}.   
Our current SPH code does not allow for particles to share their metal content with one another, and we plan to implement and explore the effects of metal diffusion in the future. 
Finally, as a comparison to the $\htwo$-SF model, we also plan to develop a turbulence-based SF model and explore the differences between the two approaches to star formation.

\section*{Acknowledgements}
We are grateful for the helpful and insightful discussions with Mark Krumholz and Brant Robertson.
We thank Volker Springel for providing us with the original version of GADGET-3. 
This work is supported in part by the National Science Foundation (NSF) grant AST-0807491, 
National Aeronautics and Space Administration (NASA) under Grant/Cooperative 
Agreement No. NNX08AE57A issued by the Nevada NASA EPSCoR program, and 
the President's Infrastructure Award from UNLV. 
Support for Program number HST-AR-12143.01-A was provided by NASA through a grant from the Space Telescope Science Institute, which is operated by the Association of Universities for Research in Astronomy, Incorporated, under NASA contract NAS5-26555.
This research is also supported by the NSF through the TeraGrid resources provided by the Texas Advanced Computing Center (TACC) and the National Institute for Computational Sciences (NICS). 
KN is grateful to the hospitality and the partial support from the Kavli Institute for Physics and Mathematics of the Universe (IPMU), University of Tokyo, the Aspen Center for Physics, and the NSF grant No. 1066293.

\end{document}